\documentclass[a4paper,11pt]{article}
\pdfoutput=1 

\usepackage{jcappub} 

\usepackage[varg]{txfonts}
\usepackage[utf8]{inputenc}

\usepackage{longtable}
\usepackage{epsfig,caption}
\usepackage{color}

\usepackage{lineno} 

\newcommand{\lya}{Ly\ensuremath{\alpha}}%

\title{\bf  Neutrino masses and cosmology with  Lyman-alpha forest  power spectrum }

\author[a]{Nathalie Palanque-Delabrouille,}
\author[a]{Christophe Y\`eche,}
\author[a]{Julien Baur,}
\author[a]{Christophe Magneville,}
\author[b]{Graziano Rossi,}
\author[c,d,e]{Julien Lesgourgues,}
\author[a,f]{Arnaud Borde,}
\author[a]{Etienne Burtin,}
\author[a]{Jean-Marc LeGoff,}
\author[a]{James Rich,}
\author[g,h]{Matteo Viel,}
\author[i]{David Weinberg}

\emailAdd{nathalie.palanque-delabrouille@cea.fr, christophe.yeche@cea.fr, julien.baur@cea.fr, christophe.magneville@cea.fr, graziano@kias.re.kr,  Julien.Lesgourgues@cern.ch}

\affiliation[a]{CEA, Centre de Saclay, IRFU/SPP,  F-91191 Gif-sur-Yvette, France}
\affiliation[b]{Department of Astronomy and Space Science, Sejong University, Seoul, 143-747, Korea}
\affiliation[c]{Institut de Th\'eorie des Ph\'enom\`enes Physiques, \'Ecole Polytechnique F\'ed\'erale de Lausanne, CH-1015, Lausanne, Switzerland}
\affiliation[d]{CERN, Theory Division, CH-1211 Geneva 23, Switzerland}
\affiliation[e]{LAPTh, Univ. de Savoie, CNRS, B.P.110, Annecy-le-Vieux F-74941, France}
\affiliation[f]{DGA, 7 rue des Mathurins, 92221 Bagneux cedex, France}
\affiliation[g]{INAF, Osservatorio Astronomico di Trieste, Via G. B. Tiepolo 11, 34131 Trieste, Italy}
\affiliation[h]{INFN/National Institute for Nuclear Physics, Via Valerio 2, I-34127 Trieste, Italy}
\affiliation[i]{Department of Physics and Center for Cosmology and Astro-Particle Physics, Ohio State University, Columbus, OH 43210, USA}

\date{Received xx; accepted xx}

\abstract{ 
We present constraints on neutrino masses, the primordial fluctuation spectrum from inflation, and other parameters of the $\Lambda$CDM model, using the
one-dimensional Ly$\alpha$-forest power spectrum measured by \citet{Palanque-Delabrouille2013} 
from the Baryon
Oscillation Spectroscopic Survey (BOSS) of the Sloan Digital Sky Survey
(SDSS-III), complemented by Planck 2015
cosmic microwave background (CMB) data and other cosmological probes.
This paper improves on the previous analysis by \citet{Palanque2015a} by using
a more powerful set of calibrating hydrodynamical simulations that reduces uncertainties
associated with resolution and box size, by adopting a more flexible set of nuisance parameters for describing the evolution of the intergalactic medium, by including additional freedom to account for systematic uncertainties, and
by using Planck 2015 constraints in place of Planck 2013. 

Fitting Ly$\alpha$ data alone leads to cosmological parameters in excellent agreement with the
values derived independently from CMB data, except for a weak tension on the scalar index $n_s$. Combining BOSS Ly$\alpha$ with Planck CMB constrains the
sum of neutrino masses to $\sum m_\nu < 0.12$~eV (95\% C.L.) including all identified systematic uncertainties, tighter than our previous limit (0.15 eV)
and more robust. Adding Ly$\alpha$ data to CMB data reduces the uncertainties on the optical depth to reionization $\tau$, through the correlation of $\tau$ with $\sigma_8$. Similarly, correlations between cosmological parameters help in constraining the tensor-to-scalar ratio of primordial fluctuations $r$.
The tension on $n_s$  can be accommodated by allowing for a running ${\mathrm d}n_s/{\mathrm d}\ln k$. Allowing running as a free parameter in the fits does not change the limit on $\sum m_\nu$. We discuss possible interpretations of these results in the context of slow-roll inflation.

}

\begin{document}
\maketitle
\flushbottom

\section{Introduction}
\label{sec:intro}
The flux power spectrum of the Lyman-$\alpha$ (Ly$\alpha$) forest in quasar absorption spectra is a powerful tool to study clustering in the  Universe, at redshifts  $\sim 2-4$. Compared to a model derived from a set of dedicated hydrodynamical simulations, the Ly$\alpha$-flux power spectrum can provide valuable information on the formation of structures and their evolution. Furthermore, by probing scales down to a few Mpc, the 1D flux power spectrum is also sensitive to neutrino masses through the suppression of power on small scales that neutrinos induce because 
they become non-relativistic at small redshift and they therefore free-stream during most of the history of structure formation.
We here use the 1D Ly$\alpha$  flux power spectrum measured with the DR9 release of BOSS quasar data, and a grid of 36 hydrodynamical simulations having a resolution equivalent to $3\times 3072^3$ particles in a $(100~h^{-1}~{\rm Mpc})^3$ box, to constrain both cosmology and the sum of the neutrino masses $\sum m_\nu$.

Cosmic Microwave Background (CMB) can also constrain $\sum m_\nu$. 
In the standard thermal history of the Universe, massless neutrinos have an average momentum corresponding to $3.15 T_\nu \sim 0.58$~eV at the epoch of last scattering. For $\sum m_\nu > 3\times 0.58 = 1.7~{\rm eV}$, the neutrinos are still relativistic at recombination, and have no significant impact on the primary CMB anisotropies. However, for any mass value, neutrinos leave a signature on the CMB angular power spectrum through the integrated Sachs-Wolf effect and through lensing~\cite{Hou2014,lesgourgues2013neutrino}. The latest limit on $\sum m_\nu$ from CMB data alone is at the level of 0.7~eV~\cite{Planck2015}.

\begin{figure}[htbp]
\begin{center}
\epsfig{figure= 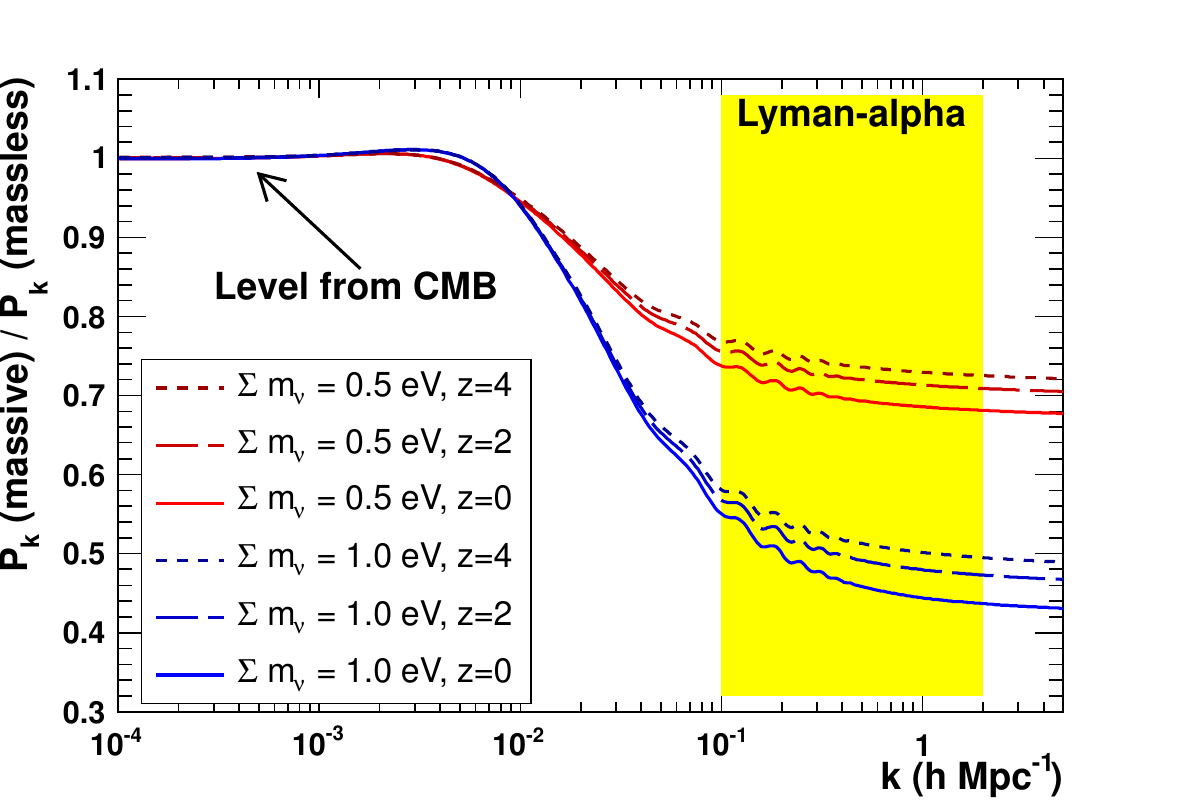,width = .8\textwidth}
\caption{\it Linear theory prediction for the matter power spectra with massive neutrinos, normalized to the corresponding massless neutrino case. The grey zone delimits  the range of $k$ covered by the 1D Ly$\alpha$ flux power spectrum from the BOSS survey.}
\label{fig:pk_ratio}
\end{center}
\end{figure}

Ly$\alpha$ data alone have  sensitivity to $\sum m_\nu$ at the level of about 1~eV due to the fact that the scales probed by Ly$\alpha$ forests are  in the region where the ratio of the power spectra for massive to massless neutrinos is quite flat  (cf. Figure~\ref{fig:pk_ratio}). However, a tight constraint on $\sum m_\nu$ can be obtained by combining  CMB data, which probe the initial power spectrum unaffected by $\sum m_\nu$, and Ly$\alpha$ data, which probe the suppressed power spectrum. Thus, Ly$\alpha$ measures the power spectrum level, defined by $\sigma_8$ and  $\Omega_m$,  CMB provides the correlations between these parameters and $\sum m_\nu$, and the joint use of these two probes significantly improves the constraint on $\sum m_\nu$ compared to what  either probe alone can achieve.

The layout of the paper is as follows. The first part of section~\ref{sec:data} presents the upgrades in the Ly$\alpha$, CMB and Baryon Acoustic Oscillation (BAO)   data sets used for this work. The second part  summarizes a number of improvements in the methodology: changes in the accounting of the uncertainties of the hydrodynamical simulations, and updates of the likelihood parameters to allow for additional freedom in the IGM model or in the instrumental systematic effects.  The main objective of section~\ref{sec:LyaAlone} is to present what Ly$\alpha$ data alone have to say about cosmology. The base model we consider is a flat $\Lambda$CDM cosmology with massive neutrinos, thereafter referred to as the base $\Lambda$CDM$\nu$ cosmology. We  start by giving the constraints measured on the five relevant parameters ($\sigma_8$, $n_s$, $\Omega_m$, $H_0$, $\sum m_\nu$), and we briefly discuss the values of the `nuisance' parameters. In section~\ref{sec:LyaCMB}, we include additional data, namely several configurations of CMB data and, occasionally, BAO measurements. We present the results obtained on the parameters of our base $\Lambda$CDM$\nu$ cosmology with various combinations of these data sets. Finally, we discuss extensions to the base $\Lambda$CDM$\nu$ cosmology. We present how Ly$\alpha$ data can contribute to constraining additional parameters through their correlations  with parameters that Ly$\alpha$ data are sensitive to;  we thus give constraints on the reionization optical depth in section~\ref{sec:tau}, and on primordial fluctuations (e.g., the running of the scalar spectral index ${\mathrm d}n_s/{\mathrm d}\ln k$ and the ratio of tensor to scalar modes $r$) in section~\ref{sec:primordial}. We discuss the small impact of the running of $n_s$ on the constraints on $\sum m_\nu$. We do not discuss here the combined Ly$\alpha$ + CMB constraint on the number of neutrino species $N_{\rm eff}$, since this  is the object of a dedicated study \cite{Rossi2015}. 

This paper refers extensively to the earlier paper~\cite{Palanque2015a} that reported the first constraints on cosmological parameters and total neutrino mass using Ly$\alpha$ data from the SDSS-III/BOSS survey. To simplify the presentation and make it easier for the reader to identify any reference to this earlier paper, we will henceforth refer to it as Paper~I. We also refer the reader to \cite{Borde2014} for a detailed description of the grid of hydrodynamical simulations used in this work, and to \cite{Rossi2014} for the implementation of neutrinos  and their impact on the 1D flux power spectrum. Definitions of the most relevant symbols used in this paper can be found in Tables~\ref{tab:astroparam} and~\ref{tab:cosmoparam}.

\begin{table}[htbp]
\caption{Definition of astrophysical parameters}
\begin{center}
\begin{tabular}{p{2.9cm}ll}
\hline
\hline
Parameter& Definition \\
\hline
$\delta=\rho /  \left\langle \rho \right\rangle$  \dotfill  & Normalized baryonic density $\rho$ of IGM\\ 
$T$ \dotfill & Temperature of IGM modeled by $T=T_0  \cdot  \delta^{\gamma-1}$ \\
$T_0$ \dotfill & Normalization temperature of IGM at $z=3$ \\
$\gamma$ \dotfill &  Logarithmic slope of  $\delta$ dependence of IGM temperature  at $z=3$ \\
$\eta^{T_0}$ \dotfill &  Logarithmic slope of  redshift dependence of $T_0$ (different for $z<$ or $>3$) \\
$\eta^{\gamma}$  \dotfill &  Logarithmic slope of  redshift dependence of $\gamma$ \\
$A^\tau$  \dotfill &  Effective optical depth of Ly$\alpha$ absorption  at $z=3$\\
$\eta^\tau$  \dotfill &  Logarithmic slope of  redshift dependence of $A^\tau$   \\
$f_ {\rm{Si\,III}}$  \dotfill &  Fraction of Si\,III absorption relative to Ly$\alpha$ absorption\\
$f_ {\rm{Si\,II}}$  \dotfill &  Fraction of Si\,II absorption relative to Ly$\alpha$ absorption\\
\hline
\end{tabular}
\end{center}
\label{tab:astroparam}
\end{table}%

\begin{table}[htbp]
\caption{Definition of cosmological  parameters}
\begin{center}
\begin{tabular}{p{4cm}l}
\hline
\hline
Parameter&  Definition \\
\hline
$\Omega_m$ \dotfill & Matter fraction today (compared to critical density) \\
$H_0$ \dotfill &  Expansion rate today in km s$^{-1}$ Mpc$^{-1}$ \\
$\sum m_\nu$ \dotfill &  Sum of neutrino masses in eV \\
$\sigma_8$ \dotfill & RMS matter fluctuation amplitude today in linear theory \\
$\tau$ \dotfill & Optical depth to reionization\\
$z_{\rm re}$ \dotfill & Redshift where reionization fraction is 50\%\\
$A_s$ \dotfill & Scalar power spectrum amplitude\\
$n_s$ \dotfill & Scalar spectral index \\
$k_0=0.05~{\rm Mpc}^{-1}$ \dotfill & Pivot scale of CMB \\
${\mathrm d}n_s/{\mathrm d}\ln k$ \dotfill &  Running of scalar spectral index \\
$r$ \dotfill &  Tensor-to-scalar power ratio\\
$\varepsilon_1 = - \dot{H}/H^2$  \dotfill & First Hubble hierarchy parameter\\
$\varepsilon_{i+1} = - \dot{\varepsilon_i}/(H\varepsilon_i)$ \dotfill & (i+1)st Hubble hierarchy parameter\\
\hline
\end{tabular}
\end{center}
\label{tab:cosmoparam}
\end{table}%


\section{Data and methodology}
\label{sec:data}

The methodology, notation and parameters used in this paper mostly follow those described in Paper~I~\cite{Palanque2015a}. We, however, include additional data, as described in Sec.~\ref{sec:newdata}. We also present a number of updates in the methodology. We include additional parameters in the likelihood to account for all the  systematic effects we identified, whether related to  the instrument or to the simulations. We have a revised  estimate of the sampling variance, and an improved  model of the impact of the splicing technique on the simulated power spectrum. We also allow for additional freedom in the model of the intergalactic medium (IGM) temperature. All these changes are described in Sec.~\ref{sec:newmodel}.

\subsection{Changes in the data}
\label{sec:newdata}

This paper is mostly focused on the analysis of Ly$\alpha$ data to provide constraints on cosmological parameters. We also include at times additional probes, namely measurements of the cosmic microwave background and of large-scale structures at redshifts $z<1$, to improve the constraints we derive.

\subsubsection{Ly$\alpha$}
\begin{figure}[htbp]
\begin{center}
\epsfig{figure= 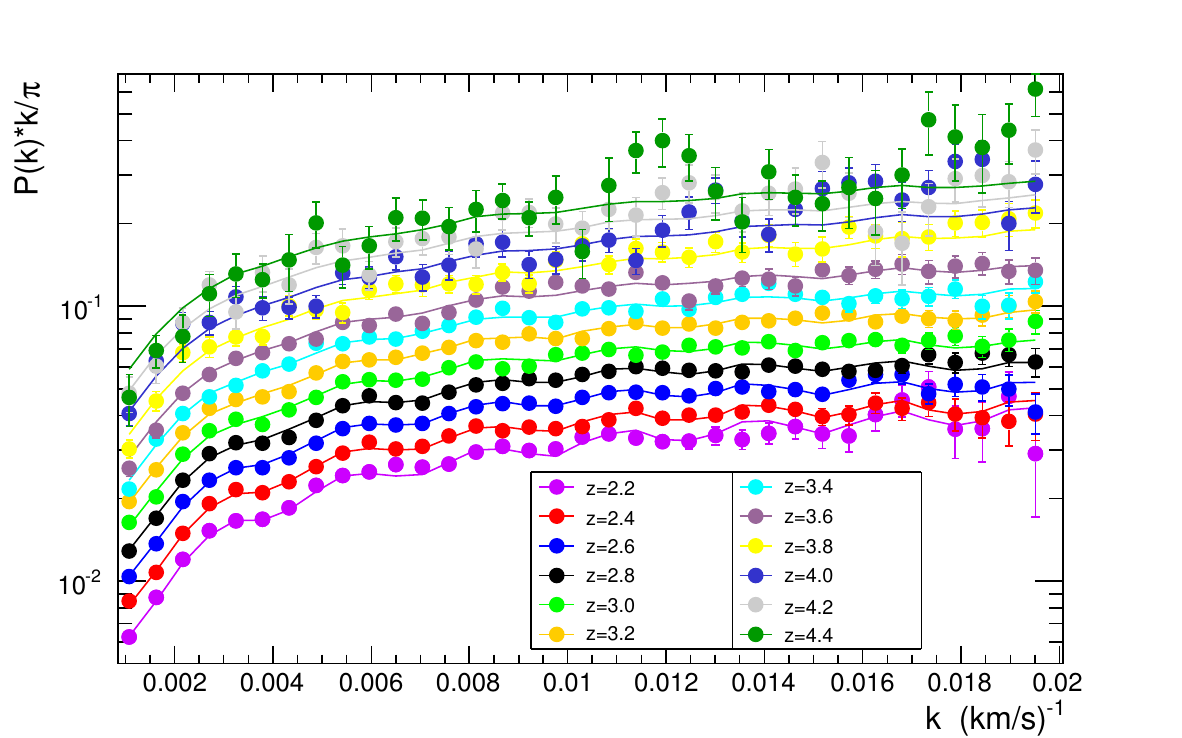,width = .8\linewidth} 
\caption{\it 1D Ly$\alpha$ forest power spectrum from the SDSS-III/BOSS DR9 data. The solid curves show the best-fit model obtained in section~\ref{sec:LyaAlone} when considering Ly$\alpha$ data alone.  The oscillations in these model predictions (and, presumably, in the measurements) arise principally from Ly$\alpha$-Si\,III correlations, which occur at a  wavelength separation $\Delta \lambda = 9.2$\AA.} 
\label{fig:P1D}
\end{center}
\end{figure}

As our large-scale structure probe, we use the 1D Ly$\alpha$-flux power spectrum measurement~\cite{Palanque-Delabrouille2013} from the first release of BOSS quasar data. The data consist of a sample of $13~821$  spectra selected from the larger sample of about $60~000$ quasar spectra of the SDSS-III/BOSS DR9 \cite{Ahn2012, Dawson2012, Eisenstein2011, Gunn2006, Ross2012,Smee2013} on the basis of their high quality, high signal-to-noise ratio and good spectral resolution ($<85\,\rm km~s^{-1}$ on average over a quasar forest).  In Paper~I, we chose  to focus the analysis on the first ten redshift bins, spanning the range $2.1<z<4.1$. With the improvement on our model of the IGM temperature, however, we are now able to use all 12 redshift bins, thus extending the redshift coverage of the analysis to $2.1<z<4.5$ as shown on Fig.~\ref{fig:P1D}. We thus now do the analysis on 420 Ly$\alpha$ data points, consisting of 12 redshift bins and 35 $k$ bins. 

\subsubsection{Cosmic microwave background}\label{sec:cmb}
In Paper~I, the cosmic microwave background (CMB) data and results we used were described in the March 2013 Planck cosmological parameters paper \cite{PlanckCollaboration2013}. When combining with CMB data, we now use the more recent  cosmological results obtained from the full Planck mission and presented in \cite{Planck2015}. We consider several subsets of Planck data. The base configuration, denoted `TT+lowP' as in  \cite{Planck2015}, uses the TT spectra at low and high multipoles and the polarization information up to multipoles $\ell=29$ (`lowP'). We also use at times the configuration based on TT, TE and EE spectra, along with the low-multipole polarization, denoted `TT+TE+EE+lowP'.  When studying inflation parameters, we will also use  the BICEP2/keck Array-Planck joint analysis~\cite{BKP2015}, which combines the high sensitivity B-mode maps from BICEP2 and Keck Array with the Planck maps at higher frequencies where dust emission dominates. This set will be denoted `BKP'.

\subsubsection{Baryon acoustic oscillations}
Finally, we  occasionally combine CMB data with measurements of the BAO scale by 6dFGS~\cite{Beutler2011}, SDSS main galaxy sample~\cite{Ross2014}, BOSS-LOWZ~\cite{Anderson2014} and CMASS-DR11~\cite{Anderson2014}. Theses measurements are henceforth globally denoted 'BAO'. The additional constraints that these measurement provide on cosmological parameters are included in the present work with their full correlation with CMB data. 
Both CMB and BAO constraints are taken from  the  Markov Chains publicly available through the official Planck Legacy Archive at http://pla.esac.esa.int. 

\subsection{Changes in the methodology and models}
\label{sec:newmodel}

We interpret the 1D Ly$\alpha$-flux power spectrum using  a likelihood built around three categories of parameters which are floated in the minimization procedure.  The first category describes the cosmological model in the simplest case of $\Lambda$CDM assuming a flat Universe. The second category models the astrophysics within the IGM, and the relationship between the gas temperature and its density.  
The  purpose of the third category is  to describe the imperfections of our measurement of the 1D power spectrum. This likelihood allows us to compare the measurement  to the power spectrum predicted from hydrodynamical simulations. The changes in the simulation model or in the likelihood compared to Paper~I are described below. 

\subsubsection{Sample variance}
\begin{figure}[p]
\begin{center}
\epsfig{figure= 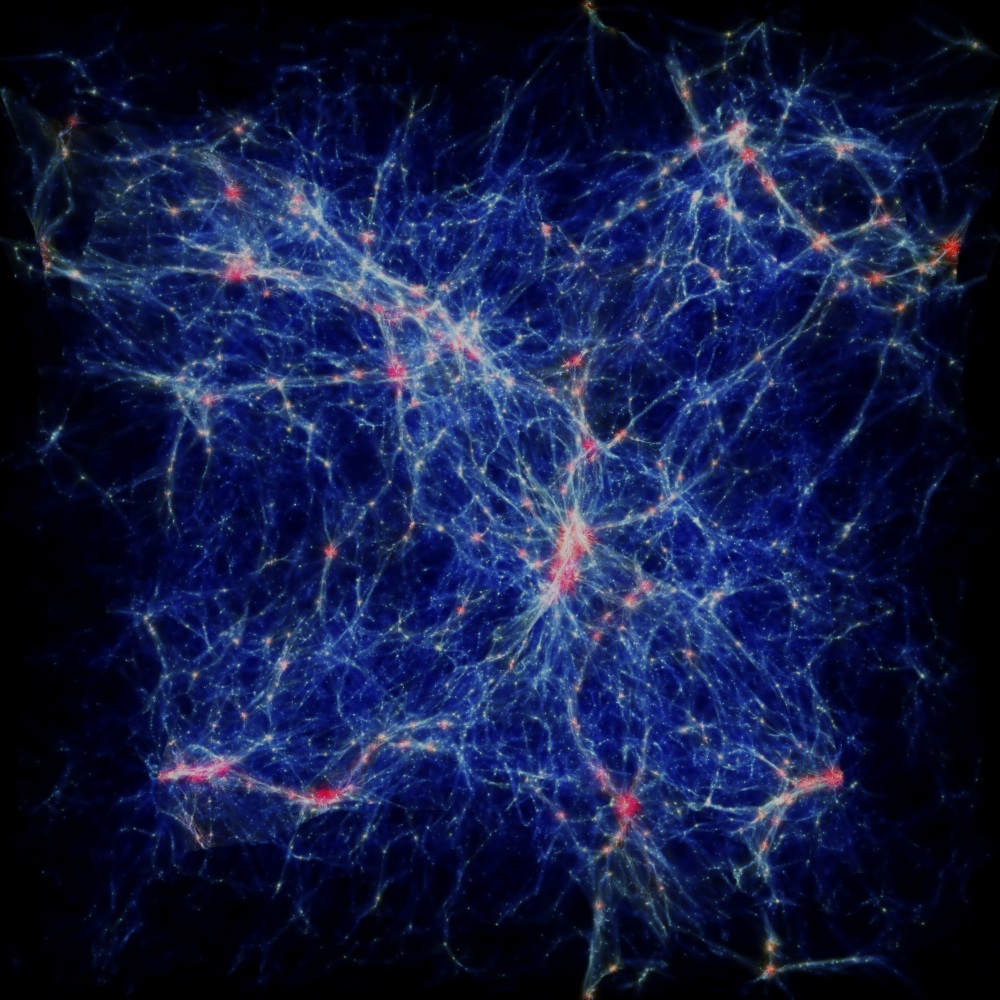,width = .45\linewidth}\hfill
\epsfig{figure= 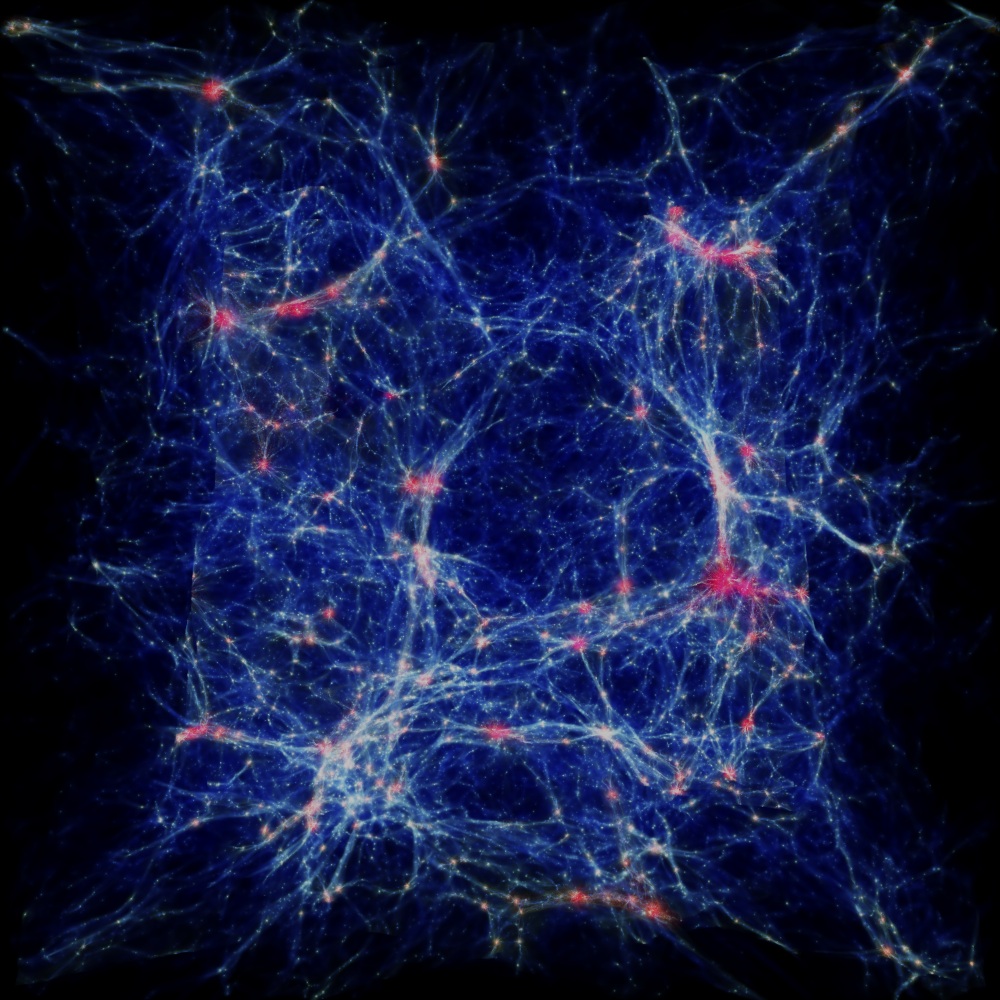,width = .45\linewidth}
\epsfig{figure= 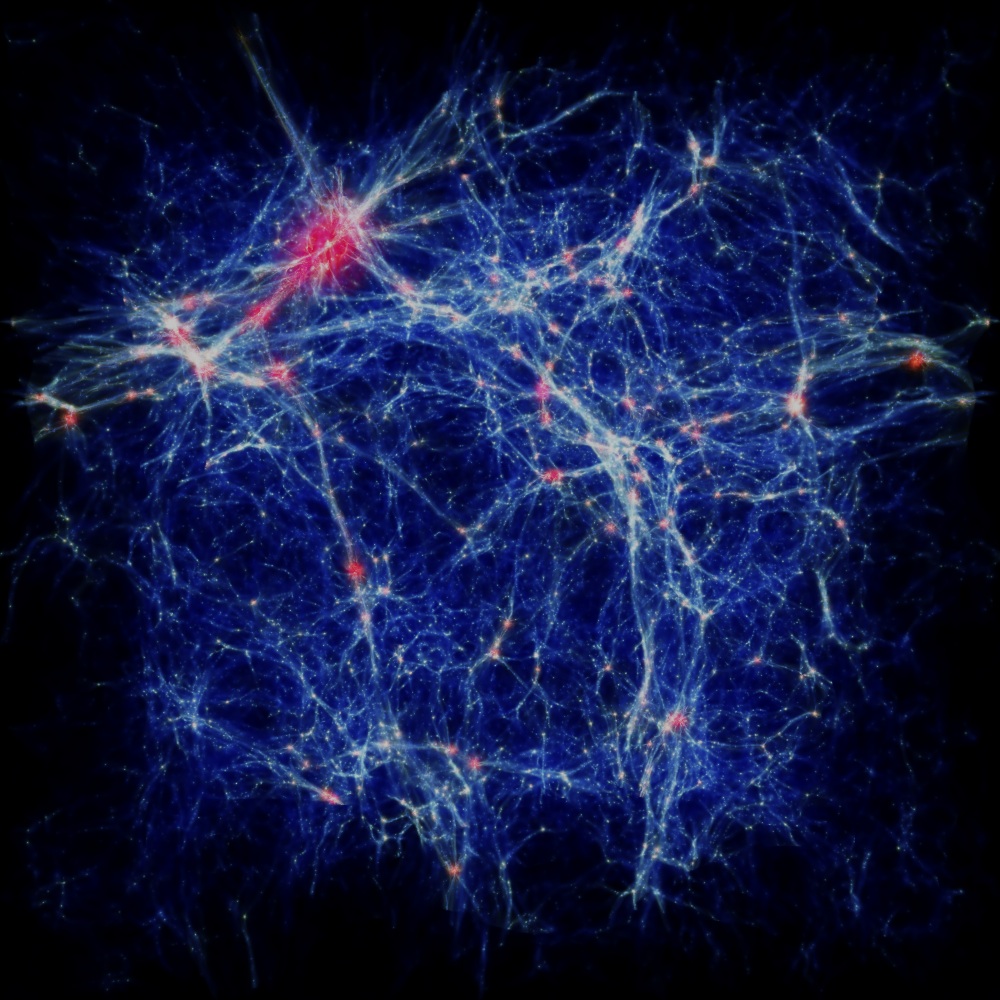,width = .45\linewidth}\hfill
\epsfig{figure= 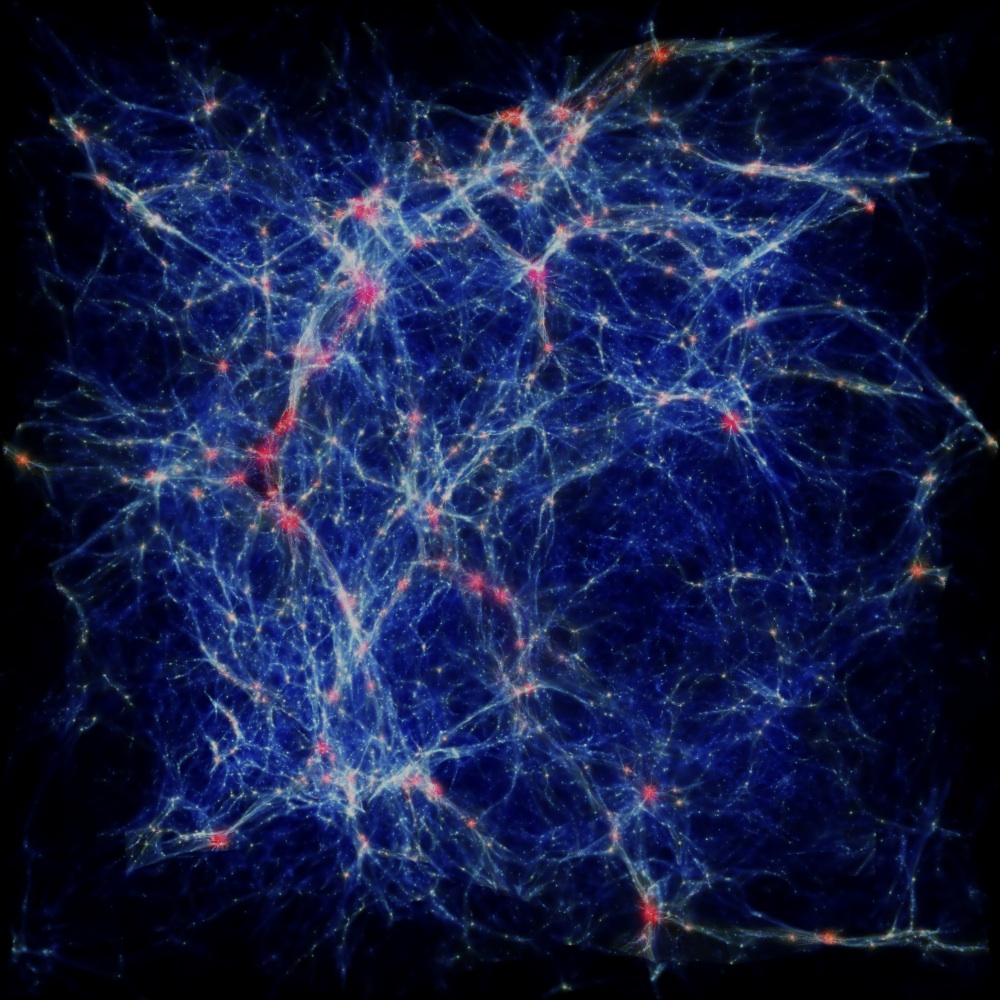,width = .45\linewidth}
\epsfig{figure= 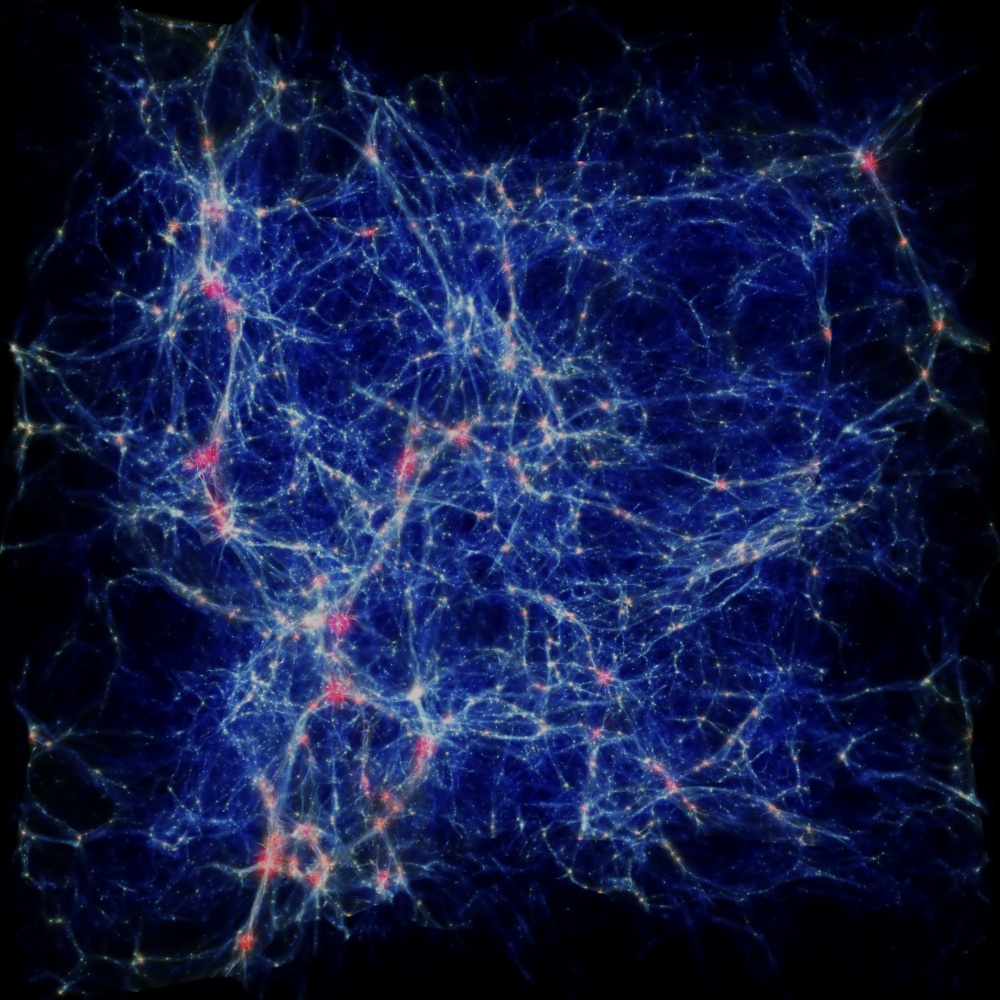,width = .45\linewidth}\hfill
\epsfig{figure= 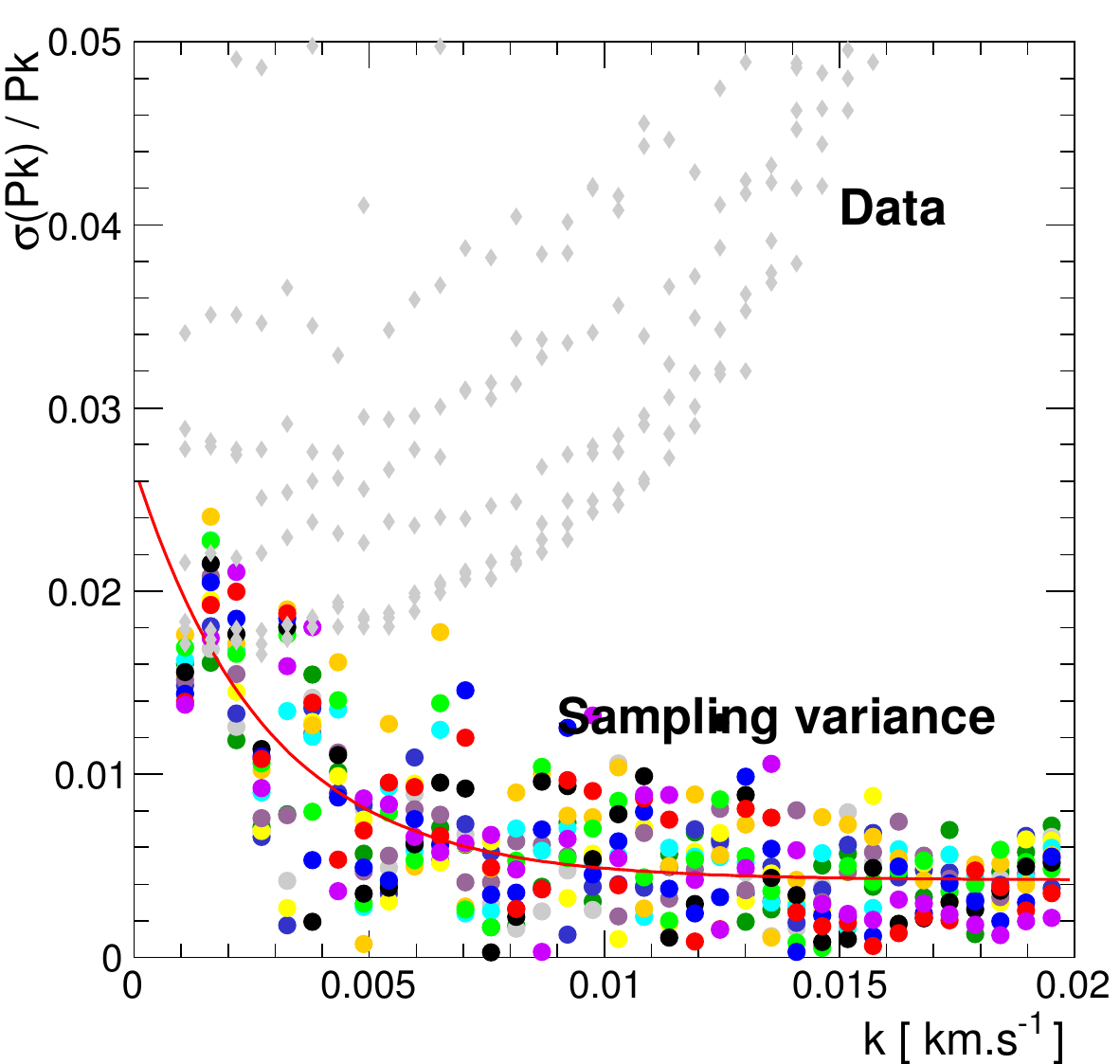,width = .45\linewidth}
\caption{\it  Visualization of the baryonic gas  at $z=2.2$  for five simulations run with identical parameters but different random seeds to compute the initial conditions.  Simulations are using $2\times 768^3$ particles in a $(100~h^{-1}~Mpc)^3$ box. Color represents  gas temperature  (from blue to red)
and density is mapped to intensity. The  plot at the bottom right illustrates the  relative total uncertainty on the data power spectrum (grey) and the level of sampling variance (colored dots for each redshift bin, same color-code as Fig.~\ref{fig:P1D}).}
\label{fig:cosmicvar}
\end{center}
\end{figure}

A sample variance is expected on large scales since the size of the simulation volume is similar to the largest modes measured. We improved our estimate of its contribution to the simulation uncertainties by computing, for each mode, the variance of the difference from  average of  the 1D Ly$\alpha$-flux power spectrum for  five simulations run with exactly the same cosmological and astrophysical parameters but  different random seeds to initiate the distribution of particles (cf. Fig.~\ref{fig:cosmicvar}). As expected, this test shows an excess of variance at small $k$, compared to the uncertainty measured within each run, which we model by a function of the form $(a + b\exp(-ck))^2$ where $k$ is the wavenumber, $a=0.004$, $b=0.023$ and $c=-356.6$. This additional variance is added in quadrature to the simulation statistical variance.

\subsubsection{Splicing}\label{sec:splicing}
\begin{figure}[htbp]
\begin{center}
\epsfig{figure= 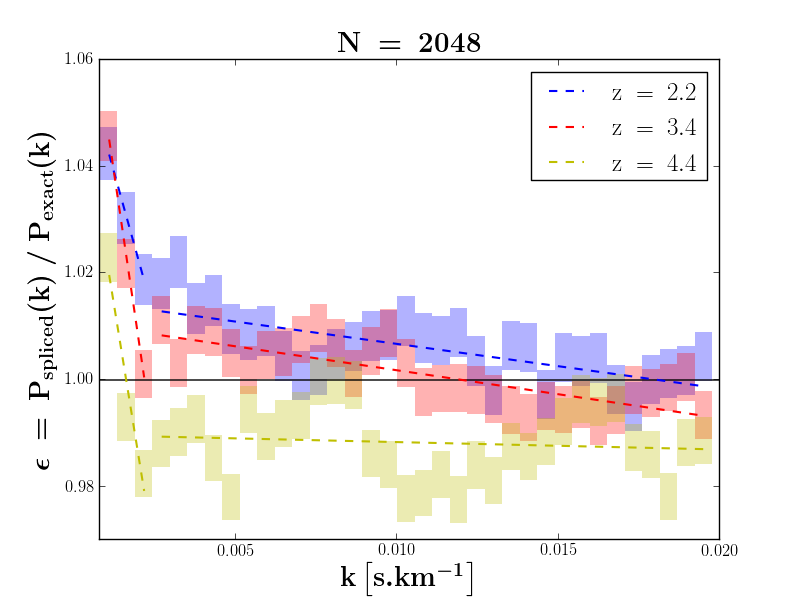,width = 12.0cm} 
\caption{\it  Residuals of the spliced to the exact power spectrum for z = 2.2,  3.4 and z = 4.4 and broken-line fit to each redshift bin individually. The break near $k\sim 0.02~{ s~km^{-1}}$ is due to the change of regime in the splicing technique.} 
\label{fig:splicing} 
\end{center}
\end{figure}

The 1D power spectrum is computed using a set of three simulations of different mass resolutions ($192^3$ or $768^3$ particles per species) and sizes (boxes 25 or 100~$h^{-1}$~Mpc on a side). A splicing technique \cite{McDonald2003}  is then used to combine the power spectra and provide the equivalent of a power spectrum measured from a single simulation spanning a total volume of $(100~h^{-1}~{\rm Mpc})^3$ with $3072^3$ particles per species. We study the accuracy of the method  by comparing a spliced power spectrum $P_{\rm spliced}$ built from a simulation set with $128^3$ or $512^3$ particles per species in a  box of side length 25 or 100~$h^{-1}$~Mpc, with the corresponding `exact' power spectrum $P_{\rm exact}$ obtained from a simulation with $2048^3$ particles per species in a box of side length 100~$h^{-1}$~Mpc.\footnote{The  `exact' simulation with $2048^3$ particles per species and $(100~h^{-1}~{\rm Mpc})^3$ box volume  required 400~khrs split among 8192 cpus for the hydrodynamical part, and a total of 300~khrs split over 128 cpus each with 4~Gb of memory for the extraction of the 100 thousand lines of sight.} The residuals $P_{\rm spliced} / P_{\rm exact}$ are shown in figure~\ref{fig:splicing}. They are modeled  by a broken line, with pivot point at the mode $k_p$ corresponding to the change of regime  from constant  to $k$-dependent resolution correction in the splicing technique (cf. details in \cite{Borde2014}). The vertical offset at $k_p$ is let free, the slope at $k<k_p$ is fixed to $-21$ or $-27~{\rm km.s^{-1}}$ for $z<3.5$ or $z>3.5$ respectively, and the slope at $k>k_p$ is let free. We allow for a small redshift dependence in the splicing correction (decreasing residual offset at the pivot point, and increasing slope in the $k>k_p$ region as $z$ increases). This model is an improvement over Paper~I where the accuracy of the splicing technique was estimated from less resolved simulations and modeled by a single redshift-independent linear function of $k$ over all modes.

\subsubsection{Model of IGM temperature}

In Paper~I, the relation between the density and the temperature of the IGM was derived directly from the simulations described in \cite{Borde2014}. Two variables $T_0$ and $\gamma$ parametrize this relation according to $T=T_0  \cdot  \delta^{\gamma-1}$ where $\delta$ is the normalized density $\rho /  \left\langle \rho \right\rangle$. For a given simulation, we measured $T_0$ and $\gamma$  by building, for each  redshift $z$,  the $ T- \rho$ diagram. In the low density and low  temperature region where the IGM lies, a linear fit of $\ln(T)$ as a function of $\ln(\delta)$ allowed us to determine  $T_0(z)$ and $\gamma(z)$. The parameter $\gamma(z)$  is monotonically and smoothly decreasing with redshift,  whereas  $T_0(z)$  presents two regimes with a break at $z\sim 3$. The latter distribution follows notably well the measurements of \cite{Becker2011}. As a consequence, in Paper~I, we  fixed the redshift dependences to that measured in the simulations, and we let free two global parameters $T_0$ and $\gamma$.

In this new analysis, we release the shapes of  $T_0(z)$ and $\gamma(z)$. Their evolution with redshift is  modeled with power laws. For $\gamma$ we introduce two parameters: $\gamma\,(z=3)$, the value of $\gamma$  at  $z=3$,  and the exponent $\eta^\gamma$ of the relation $\gamma (z) =  \gamma\,(z=3)\cdot [(1+z)/4 ]^{\eta^\gamma}$. For the temperature, we use three parameters: $T_0\,(z=3)$, the value of the temperature at  $z=3$, and two slopes  $\eta^{T_0}\,(z<3)$ and $\eta^{T_0}\;(z>3)$ to take into account the break at redshift $z\sim3$. 

In summary five parameters, $T_0 \;(z=3)$, $\eta^{T_0}\;(z<3)$,  $\eta^{T_0}\;(z>3)$, $\gamma\;(z=3)$ and  $\eta^\gamma$  are now floated in the fit to model the density and the temperature of the IGM.

\subsubsection{Reionization history}
The hydrogen reionization history will alter the pressure smoothing scale of gas in the IGM, particularly at redshifts
approaching the tail-end of the reionization at $z_{\rm re}\sim 6$~\citep{Gnedin1998a}. Increasing the redshift of reionization allows more time for pressure to suppress small-scale structures. In the range of scales probed by the BOSS Ly$\alpha$ data,  Fig.~13 of \citep{McDonald2005} shows that an increase of the redshift of reionization from $z_{\rm re}=7$ to 17 suppresses the Ly$\alpha$ flux power spectrum at the largest modes ($k \sim 0.02~{\rm s.km^{-1}}$) by about 1\% at $z=2.1$ and 5\% at $z=4.0$. This range for  $z_{\rm re}$ is extreme considering recents limits from Planck temperature plus low-$\ell$ polarization data, which give $z_{\rm re}=9.9^{+1.8}_{-1.6}$ (68\% CL). Current uncertainties on $z_{\rm re}$ therefore translate into shifts of a few percent at most  on the flux power spectrum, dominantly at high redshift and on small scales. 

In this work, we  introduced two nuisance parameters that absorb the effects of   different reionization histories.
First, we have a nuisance parameter that accounts for the  error caused by the limited simulation resolution: the slope at $k>k_p$ of the splicing correction described in Sec.~\ref{sec:splicing}.  
As was already noted by~\citet{McDonald2005}, such a nuisance parameter plays a dual role, since it also absorbs uncertainty related to the reionization optical depth (or, equivalently, to the smoothing of the baryon field). The allowed variation range of the slope ($\pm 8\%$ at $1\sigma$ for the largest modes) fully encompasses the maximum range of variation caused by different $z_{\rm re}$.
Secondly, we  also introduced additional freedom compared to \cite{McDonald2005} by including a redshift-dependence of the nuisance parameter accounting for uncertainty in the spectrograph resolution (cf. Sec.~\ref{sec:techn_nuisance}). This correction is also only impacting small scales, with a range of about 1\% at $1~\sigma$ for the redshift coverage of the BOSS data. 

An additional nuisance parameter accounting for different reionization histories would be  degenerate with the nuisance parameters dedicated to the slope of the splicing  correction and to the redshift-dependence of the spectrograph resolution.

\subsubsection{Technical nuisance parameters}\label{sec:techn_nuisance}
In Paper~I, we identified several sources of systematics, and we tested their impact  on the cosmological  and neutrino mass bounds we derived. In the present work, these systematics are all included as nuisance parameters free to vary in the fit. On the technical side, we varied a parameter  to account for a possible redshift-dependence of the correction to the spectrograph resolution, such that the resulting correction to the spectrograph resolution is now the  multiplicative factor $ {\cal C}_{\rm reso}(k) =  \exp[ - k^2\cdot (\alpha_{\rm reso} + \beta_{\rm reso}\,(z-3))]$. Both $\alpha_{\rm reso}$ and $ \beta_{\rm reso}$ are allowed to vary around 0 with a Gaussian constraint $\sigma = (5\,{\rm km~s^{-1}})^2$.

\subsubsection{Simulation nuisance parameters} \label{sec:nuisanceIGM}
We include  parameters to allow for additional freedom in our model of the IGM. We follow the study described in section 5.2 ({\em impact of simplifying hypotheses}) of Paper~I, which we refer to for details on the analytical form of each correction. The multiplicative corrections are normalized to 1 for an impact identical to the systematic effect of Paper~I. Since the cases we had assumed were quite extreme, we  here allow the corresponding  parameter to vary around 0 with a Gaussian constraint $\sigma=0.3$.  

Although we remove all quasars flagged as having damped Lyman alpha (DLA)  or detectable Lyman limit systems in their forest, we introduce a multiplicative factor $1 - [1/(15000.0\;k-8.9) + 0.018]\cdot 0.2\cdot \alpha_{\rm DLA}$ to account for  a possible remaining contribution of high-density absorbers in the quasar spectra. This form is motivated by the study led by \citet{McDonald2005}, and $\alpha_{\rm DLA}$ is free to vary in the fit.

We model AGN and SN feedbacks  by multiplicative factors of the form  ${\cal C}_{{\rm feedback},i}(k) = (\alpha_i(z) + \beta_i(z)\;k)\cdot \alpha_{{\rm feedback}, i}$, where $i$ stands for either AGN or SN. The coefficients $\alpha_i(z)$ and $\beta_i(z)$ are derived from \cite{Viel12}. Over our range of interest, $0.001<k<0.02$, the slope of the power spectrum can be reduced by 6\% at most at $z<2.5$ and 2\% for $z>3.5$ when adding  AGN feedback, while it is increased by 5\% and 1\% for the same redshift bins respectively in the case of SN feedback. 

Fluctuations in the intensity of the ionizing background (also referred to as UV fluctuations) are accounted for by an additive  correction proportional to the measured power spectrum at the pivot wavenumber $k_p=0.009\; {\rm s~km^{-1}}$, as motivated by the study of \citet{Satya2014}. The normalization factor is a floating parameter.  The correction is k-independent but evolves with redshift proportionally to the power spectrum. 

\subsection{Impact of neutrino mass hierarchy}

Cosmology is often said to be mainly sensitive to neutrino masses through their contribution to the energy density, i.e., through $\Omega_\nu$, and therefore only through their total  mass $\sum m_\nu$. While this is true to very good approximation for the CMB spectrum, the impact of individual neutrino masses on large scale structures is a little more subtle. Although  the main effect remains that of the total mass, matter power spectrum measurements have some  sensitivity to individual masses because of two effects:  (i) the detailed evolution of the background density close to the time of the non-relativistic transition of each species depends on individual masses, and (ii) so does the free-streaming scale of each species. When the individual masses are varied for the same total mass, these two effects lead respectively to different amplitude and shape in the small-scale matter power spectrum~ \cite{Lesgourgues:2006nd,lesgourgues2013neutrino}.

Throughout this paper, we assume the three neutrino species to share a common mass equal to $\sum m_\nu / 3$. Neutrino oscillation measurements, however, have shown that the three neutrino species have slightly different masses. According to the compilation of~\cite{Capozzi:2013csa} from the combination of atmospheric, solar, reactor and accelerator neutrino experiments,  the masses verify 
$\delta m^2 = 7.54\pm 0.24 \times 10^{-5} ~{\rm eV^2}$ and $\Delta m^2 = 2.43\pm 0.06\times 10^{-3}~{\rm eV^2}$ (using the authors' formalism). The masses can follow a normal hierarchy (NH) with two light states  and a heavier one, in which case the minimum total mass is $\sum m_\nu=0.06$~eV. In case of inverted hierarchy (IH), the two heavy states are split by $\delta m^2$  and the lighter one is separated from the other two by $\Delta m^2$;  the   minimum total mass is then 0.10~eV.

As the bounds on $\sum m_\nu$ get closer to the 0.10~eV upper limit where we can distinguish between normal and inverted  hierarchies, it becomes increasingly critical to test the impact of mass hierarchy on the derived 1D flux power spectrum. We  therefore ran 3 simulations with identical cosmological and astrophysical parameters, as well as same total neutrino mass $\sum m_\nu =  0.10$~eV. The  difference between the three simulations resides in the assumed neutrino mass hierarchy: the first one is produced with degenerate neutrino masses (DM), the second one with normal hierarchy (NH), and the third one with inverted hierarchy (IH). In the latter two cases, the mass of each species is determined according to the squared mass differences of~\cite{Capozzi:2013csa}.
The masses of the individual species in each case are given in Tab.~\ref{tab:nuhierarchy}.
\begin{table}[htb]
\caption{\it Masses of each neutrino species (in eV) for the three configurations of  mass hierarchy considered in the text. }
\begin{center}
\begin{tabular}{lcccc}
\hline
Hierarchy&$m_1$&$m_2$&$m_3$&$\sum m_\nu$ \\
\hline \\[-10pt]
Degenerate & 0.033&0.033&0.033&0.100\\
Normal & 0.022 & 0.024& 0.055& 0.100\\
Inverted & 0.0007&0.049& 0.050& 0.100\\
\hline
\end{tabular}
\end{center}
\label{tab:nuhierarchy}
\end{table}

\begin{figure}[htbp]
\begin{center}
\epsfig{figure= 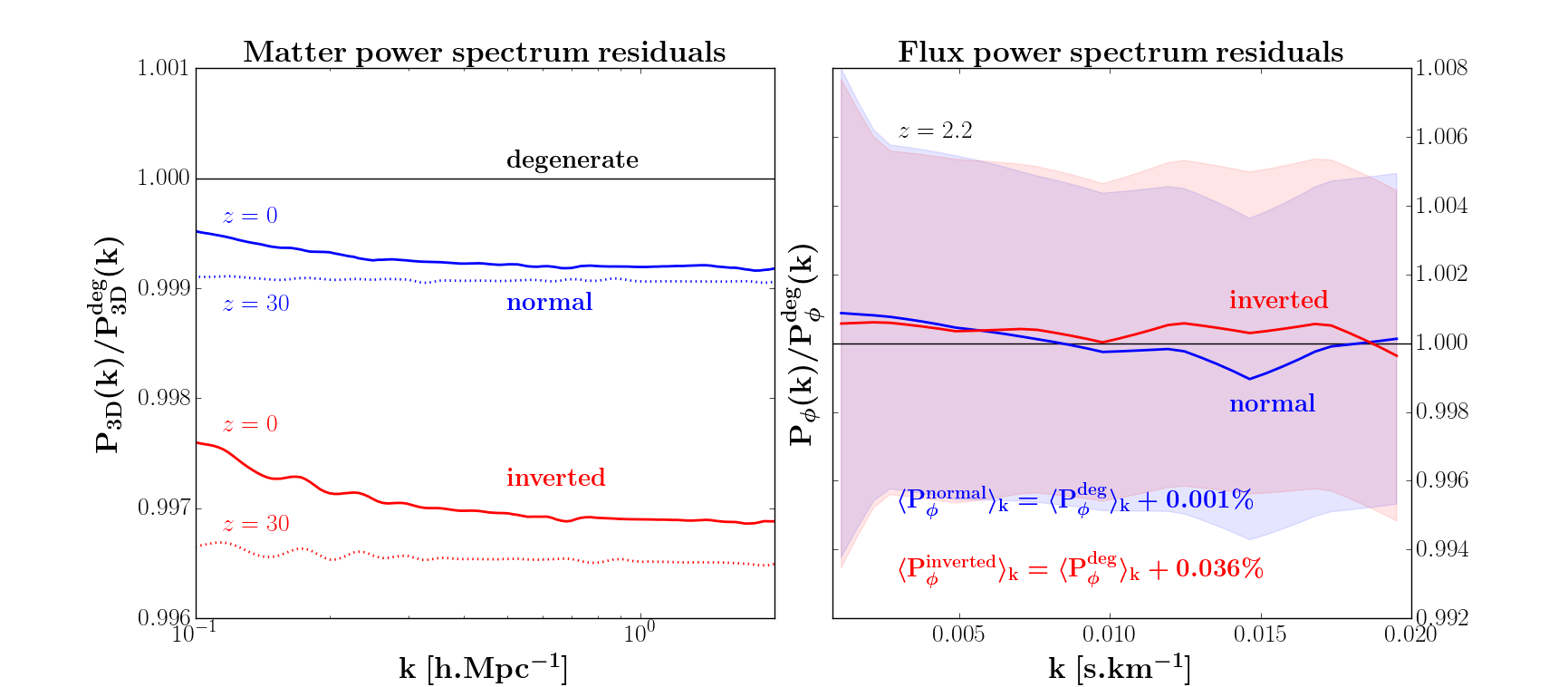,width =\linewidth} 
\caption{\it  Ratio of the 3D matter power spectra (left) and of the 1D flux power spectra (right) measured with normal (blue) or inverted (red) neutrino mass hierarchy to the power spectrum obtained assuming three degenerate-mass neutrinos, for a total neutrino mass $\sum m_\nu=0.1$~eV and identical astrophysical and cosmological parameters. The shaded area in the right plot illustrates the range of  simulation statistical uncertainties for 35 $k$-modes over the range $0.001-0.020~{\rm s~km^{-1}}$ as for  data (cf. Fig.~\ref{fig:P1D}). } 
\label{fig:nuhierarchy}
\end{center}
\end{figure}

As shown in Figure~\ref{fig:nuhierarchy} (left plot), we observe differences in the linear 3D matter power spectrum at the level of 0.1\% between normal and degenerate hierarchies, and of order 0.3\% between inverted and degenerate hierarchies, quasi independently of the scale $k$. The larger difference in the case of inverted hierarchy is explained by the fact that the total mass is essentially shared among two neutrinos instead of three for either the DM or the NH scenario. In the IH case, the lightest neutrino remains relativistic until late times (it becomes non-relativistic around $z_{\rm NR}\sim3$ for $\sum m_\nu=0.1$~eV). It thus contributes longer to the background density, thereby leading to a slower growth of cold dark matter perturbations and to a stronger overall suppression of power. In contrast, the NH scenario having three neutrinos of almost equal masses is closer to the case of three degenerate-mass neutrinos. The slope that appears at small $k$ is an excess of power for IH compared to NH or DM (and excess of power of NH compared to DM), due to the presence of two (or one, respectively) higher-mass neutrinos, causing an earlier  non-relativistic transition and thus a free-streaming damping restricted to smaller scales. Relatively  to the degenerate case, normal hierarchy (and even more so inverted hierarchy) therefore exhibits an excess of power near $k_{\rm NR}\sim 10^{-2}~h.{\rm Mpc}^{-1}$. The tail of this peak is the cause of the slope near $k= 10^{-1}~h.{\rm Mpc}^{-1}$ in Fig.~\ref{fig:nuhierarchy}. The effects of individual masses are nevertheless small, of  0.3\% at most.

The differences between hierarchies are  even smaller when comparing 1D flux power spectra. As shown in the right plot of Figure~\ref{fig:nuhierarchy}, the ratio of the 1D flux power spectrum measured for inverted or normal hierarchy to the power spectrum for degenerate masses is compatible with 1 at better than 0.05\%, more than 10 times below the level of the statistical uncertainty in the simulations. Both plots are zoomed on the scales that Ly$\alpha$ data can probe. The conversion between wavenumbers expressed in ${\rm s~km^{-1}}$ and in $h~{\rm Mpc}^{-1}$ is redshift dependent and given by the factor $H(z)/(1+z)$. For the central cosmology used in our simulation grid and at redshifts $z=30$, $4$, $3$, $2$ and $0$, the conversion factor equals 311, 126, 113 and 100~$\rm km~s^{-1}~{Mpc}^{-1}$, respectively.

This test clearly justifies the simplifying hypothesis of degenerate masses used in this work. The effects of individual neutrino masses are too small to be measured with current experiments.

\subsection{Interpretation methodology} 
In Paper~I, we have simultaneously used  Markov-Chain Monte-Carlo (MCMC) simulations  with Bayesian inference and a frequentist interpretation. The results obtained with either method were in  remarkably good agreement. In this paper, due to the increase in the number of parameters, all the results are derived with the faster frequentist approach (see~\cite{Yeche2006,PlanckCollaboration2014Freq} ). 

Our determination of the coverage intervals of unknown cosmological parameters is based on the `classical' confidence level method originally defined by \citet{neyman1937}.   We start with the likelihood ${\cal L}\bigl(x,\sigma_x;\Theta)$, for a given cosmological model defined by the $n$ cosmological, astrophysical and nuisance parameters $\Theta=(\theta_{1},\ldots,\theta_{n})$, and for data measurements $x$ with Gaussian experimental errors $\sigma_{x}$.  In the rest of this paper, we adopt a $\chi^2$ notation, which means that the following quantity is minimized:

\begin{equation}
\chi^2(x,\sigma_x;\Theta) = -2 \ln ({\cal L}(x,\sigma_x;\Theta))\,.
\label{eq:chi2}
\end{equation}

We first determine the minimum $\chi^2_0$ of $\chi^2(x,\sigma_{x};\Theta)$ leaving  all the cosmological parameters free. Then, to set a confidence level (CL) on any individual cosmological parameter $\theta_i$, we scan the variable $\theta_i$: for each fixed value of $\theta_i$, we minimize again $\chi^2(x,\sigma_{x};\Theta)$ but with $n-1$ free parameters. The $\chi^2$ difference, $\Delta \chi^2(\theta_i)$, between the new minimum and  $\chi^2_0$, allows us to compute the CL on the variable, assuming that the experimental errors are Gaussian,
\begin{equation}
{\rm CL}(\theta_i) = 1-\int_{\Delta \chi^2(\theta_i)}^{\infty}  f_{\chi^2}(t;N_{dof}) dt,
\label{Eq:CL}
\end{equation}
with
 \begin{equation}
 f_{\chi^2}(t;N_{dof})=\frac{e^{-t/2}t^{N_{dof}/2 -  1}}{\sqrt{2^{N_{dof}}} \Gamma(N_{dof}/2)}   \label{Eq:chi2}
\end{equation}
where $\Gamma$ is the Gamma function and the number of degrees of freedom $N_{dof}$
is equal to 1.
This profiling method can be easily extended to two variables. In this case, the minimizations are
performed for $n-2$ free parameters and the confidence level ${\rm CL}(\theta_i,\theta_j)$ is
derived from Eq.~\ref{Eq:CL} with $N_{dof}=2$.

In this paper we also combine the  $\chi^2$ derived from the Ly$\alpha$ likelihood with that of Planck. However, we do not directly use  the Planck  likelihoods. Instead, we use the central values and the covariance matrices available in the official 
 Planck~\footnote{\tt http://wiki.cosmos.esa.int/planckpla2015/index.php/Main\_Page} 
 repositories for the cosmological parameters  ($\sigma_8$, $n_s$ ,$\Omega_m$,  $H_0$,  $\sum m_\nu$, ${\mathrm d}n_s/{\mathrm d}\ln k$, $r$ and $\tau$). For each parameter, we assume a Gaussian CMB likelihood with asymmetric $1\sigma$ errors that we estimate on either side of the central value from the $1\sigma$ lower and upper limits, thus accounting for asymmetric contours.  We validated this strategy for some configurations by comparing it to  the MCMC approach using the full likelihood. The results obtained for the two methods are  quite similar, as already observed in Paper~I.


\section{Cosmological implications of Ly$\alpha$ data}
\subsection{$\Lambda$CDM$\nu$ cosmology  from Ly$\alpha$ data alone    }
\label{sec:LyaAlone}
The methodology upgrades presented in section~\ref{sec:data} impact the constraints we derive on the parameters of the base $\Lambda$CDM$\nu$ cosmology compared to Paper~I, even for Ly$\alpha$ data alone. In particular, the additional freedom in the model of the IGM temperature allows us to fit all 12 available redshift bins, thus covering $2.1<z<4.5$, whereas we limited the analysis to the first ten bins only ($2.1<z<4.1$) in Paper~I. 
As explained in the previous section, we have also more accurately determined the  contribution of sample variance to the simulation uncertainties. Finally, we have updated the likelihood with more parameters to describe the IGM temperature and its evolution with redshift,  to include  a possible contribution from AGN or SN feedbacks though a parameterized correction to the power spectrum,  and to better model the correction for the use of the splicing technique or the correction for uncertainty on the model of the spectrograph resolution. 

\begin{table}[htb]
\caption{\it Best-fit value and 68\% confidence levels (95\% upper bound in the case of  $\sum m_\nu$) of the parameters of the model fitted to the flux power spectrum $P(k_i,z_j)$ measured with the BOSS Ly$\alpha$ data. The Gaussian constraint on $H_0$ is defined by Planck 2013 cosmological results (column 1), Planck 2015 results (column 2) or Efstathiou's reanalysis of Cepheid data (column 3).  }
\begin{center}
\begin{tabular}{lccc}
\hline
Parameter &   (1) Ly$\alpha$  + $H_{0}^\mathrm{Gaussian}$  &   (2) Ly$\alpha$  + $H_{0}^\mathrm{Gaussian}$  &  (3) Ly$\alpha$ + $H_{0}^\mathrm{Gaussian}$  \\
& {\scriptsize ($H_0 = 67.4 \pm 1.4$)} & {\scriptsize ($H_0 = 67.3 \pm 1.0$)}  &{\scriptsize ($H_0 = 70.6 \pm 3.3$)}  \\

\hline \\[-13pt]
$\sigma_8$ &  $0.830\pm0.032$  &  $0.831\pm0.031$  & $0.831\pm0.032$ \\[2pt]
$n_s$ &  $0.939\pm0.010$ &  $0.939\pm0.010$ &  $0.932\pm0.010$ \\[2pt]
$\Omega_m$ &  $0.293\pm0.013$ &  $0.293\pm0.014$ &  $0.292\pm0.014$ \\[2pt]
$H_0$~{\scriptsize(${\rm km~s^{-1}~Mpc^{-1}}$)} &  $67.4\pm1.2$ &  $67.3\pm1.0$ & $70.2\pm1.0$ \\[2pt]
$\sum m_\nu$~{\scriptsize(eV)} & $<1.1$~{\scriptsize(95\% CL)} & $<1.1$~{\scriptsize(95\% CL)} & $<1.1$~{\scriptsize(95\% CL)} \\[2pt]
\hline \\[-13pt]
$f_ {\rm{Si\,III}}$ &  $0.0059\pm0.0004$ &  $0.0059\pm0.0004$ &  $0.0058\pm0.0004$ \\[2pt]
$f_ {\rm{Si\,II}}$ &  $0.0007\pm0.0005$ &  $0.0007\pm0.0005$&  $0.0007\pm0.0004$ \\[2pt]
$T_0\;(z=3)$ \scriptsize{(K)}& $9500 \pm 3500$ & $8900\;_{-4000}^{+3800}$& $8500\;_{-2500}^{+4400}$  \\[2pt]
$\eta^{T_0}\;(z<3)$& $-2.9\pm0.4$ & $-2.9\pm0.5$  & $-3.1\pm0.2$   \\[2pt]
$\eta^{T_0}\;(z>3)$& $-4.4\pm 1.0$  & $-4.4\pm 1.1$   & $-4.5\pm 0.5$  \\[2pt]
$\gamma$& $1.0\pm 0.2$ & $0.9 \pm 0.1$ & $1.0\pm 0.2$\\[2pt] 
$\eta^{\gamma}$ & $1.0\pm 0.2$& $0.8\pm 0.5$  & $0.8 \pm 0.2$  \\[2pt]
$A^\tau$ & $0.0026\pm0.0001$& $0.0025\pm0.0001$ & $0.0025\pm0.0001$ \\[2pt]
$\eta^\tau$ &$3.734\pm0.015$ &$3.73\pm0.02$ &$3.729\pm0.005$  \\[2pt] 
\hline\\[-11pt]
reduced $\chi^2$ & 0.99 & 0.99 & 0.99\\[2pt]
\hline
\end{tabular}
\end{center}
\label{tab:fit_Lya}
\end{table}

As in Paper~I, we impose a Gaussian constraint on the Hubble constant $H_0$. This is justified by the very weak dependence of Ly$\alpha$ data on $H_0$. Indeed, letting $H_0$ free in the fit does not significantly alter the best-fit value of any of the other parameters, but only constrains the Hubble constant to $H_0<77.0~{\rm km~s~Mpc^{-1}}$ at 95\% CL. 
In table~\ref{tab:fit_Lya}, we summarize the  results obtained on the fit to the  Ly$\alpha$ 1D flux power spectrum  for three different $H_0$ constraints: in column (1)  we impose the same constraint as in Paper~I (coming from Planck 2013 measurements~\cite{PlanckCollaboration2013}), in column (2) we use a constraint taken from the Planck 2015 TT + lowP results~\cite{Planck2015} and in column (3) we use a  constraint taken from the reanalysis of Cepheid data by Efstathiou~\cite{Efstathiou2014}. There is no notable difference in the values of any of the parameters whatever the constraint imposed on $H_0$. The resulting $\chi^2$ is not affected either ($\Delta \chi^2\ll 1$) by this constraint. The 2D constraints in the $n_s-\sigma_8$, $\sum m_\nu - \Omega_m$ and $\sum m_\nu - \sigma_8$ planes are shown as the red contours in Fig.~\ref{fig:OmegamMnuLyaPlanck}. The neutrino mass is correlated to  $\sigma_8$ (-48\%),  $n_s$ (48\%) and  $\Omega_m$ (52\%). Correlations between all other cosmological parameters have smaller amplitudes. Astrophysical parameters, in contrast, exhibit higher correlations, with $\sim 70\%$ correlation between $T_0$, $\gamma$ and $\eta^\tau$.

The fitted values of the nuisance parameters are all well within the expected range. The correction to the power spectrum due to the use of  the splicing technique is fitted with a small negative slope at $k>k_p$ and percent-level offset  at mode $k_p$,  in good agreement with  Fig.~\ref{fig:splicing}. Both values are compatible with 0 at $2\sigma$.The best-fit value of the correction to the estimate of the spectrograph resolution indicates an overestimate ranging from  3.5 to 4.0 ${\rm km~s^{-1}}$ from  lowest (mean $z=2.2$) to highest (mean $z=4.4$)  redshift  bin, compatible within error bars with no correction. The IGM nuisance parameters  are all also compatible with 0, with best-fit values indicating 4\%, 7\% and 15\% of the maximum DLA, SN feedback and UV fluctuation  effects described in Sec.~\ref{sec:nuisanceIGM}. There is no indication of any contribution from AGN feedback. The IGM temperature parameters have large error bars and are thus poorly constrained by this data set. Their values are within $1-2~\sigma$ of  typical  measurements (see, e.g. \citep{Becker2011}). As explained in Paper~I, there is also one additive noise correction parameter per redshift bin. Its value ranges from $-9\%$ to $+16\%$ with median at $-2.9\%$, and the extreme values are obtained for the redshift bins at mean redshifts of $4.0$ and 3.8, where the noise level is only $\sim10\%$ at most (e.g., for the smallest scales) of the  Ly$\alpha$ power spectrum. 

The values of the cosmological parameters shown in table~\ref{tab:fit_Lya} are consistent within $1\sigma$ with their value in Paper~I. This indicates that the changes in methodology and data set have not modified our main conclusions on $\Lambda$CDM$\nu$ cosmology. We have included additional 'nuisance' parameters, which in principle should lead to increased uncertainties on fit parameters, but in most cases these are compensated for by the additional lever arm in redshift due to the extra redshift bins $z=4.2$ and $z=4.4$ that are now included. As a result, the uncertainties on the parameters are similar or even slightly reduced compared to Paper~I. The upper bound on the sum of the neutrino masses is unchanged at $1.1$~eV (95\% CL). The best-fit value of the sum of the neutrino masses has moved up slightly to 0.41~eV, although still compatible with 0 at about $1\sigma$. The main changes occur on $\sigma_8$ ($0.3\sigma$ downward shift) and $n_s$ ($1.0\sigma$ upward shift). They are mostly due to our improved model of the splicing correction, where the broken slope dependence affects the  amplitude ($\sigma_8$) and scale dependence ($n_s$)  of the fluctuations.


\subsection{$\Lambda$CDM$\nu$ cosmology from Ly$\alpha$ data and other probes }
\label{sec:LyaCMB}
\begin{figure}[htbp]
\begin{center}
\epsfig{figure= 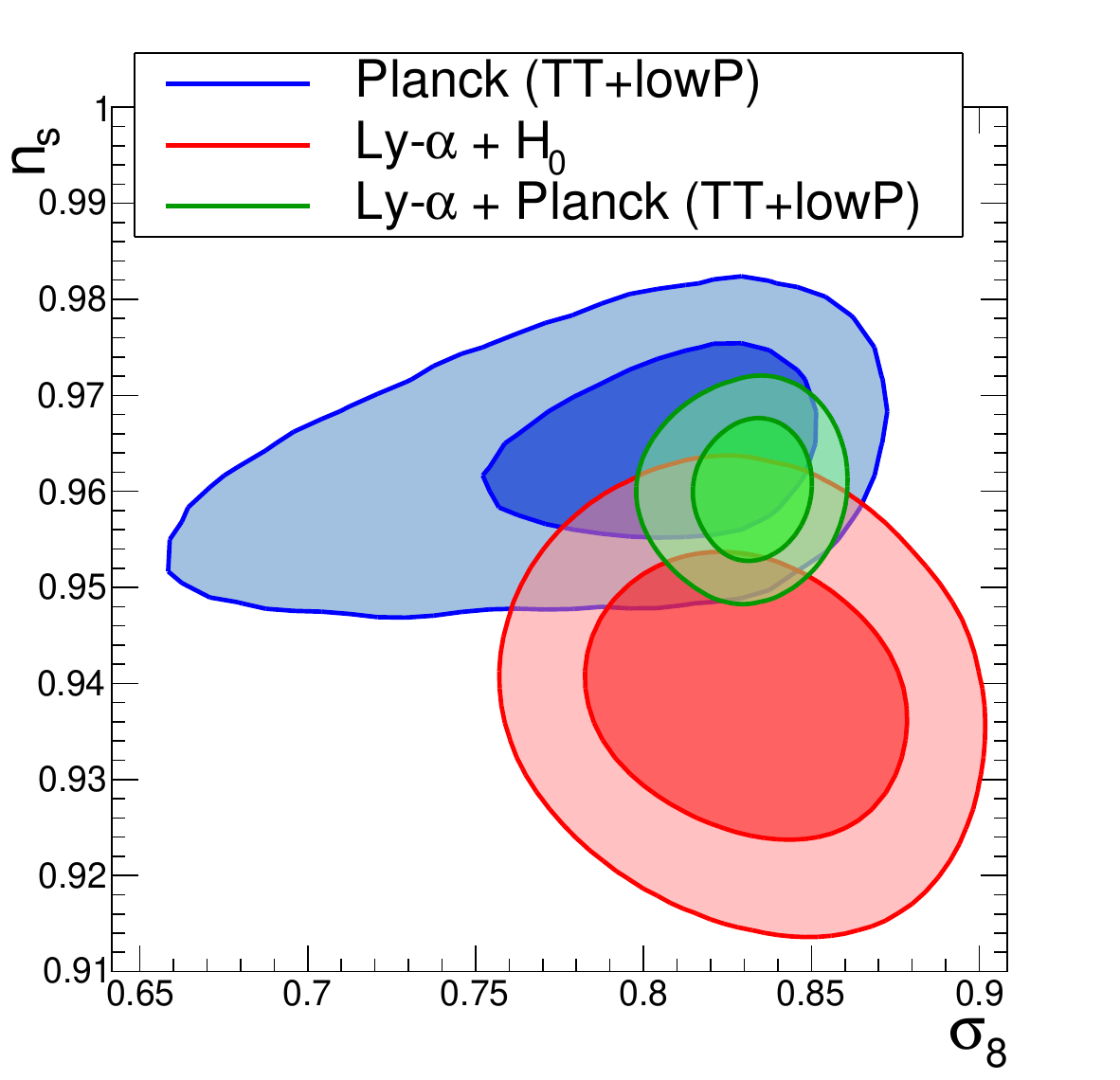,width = 5.0cm}
\epsfig{figure= 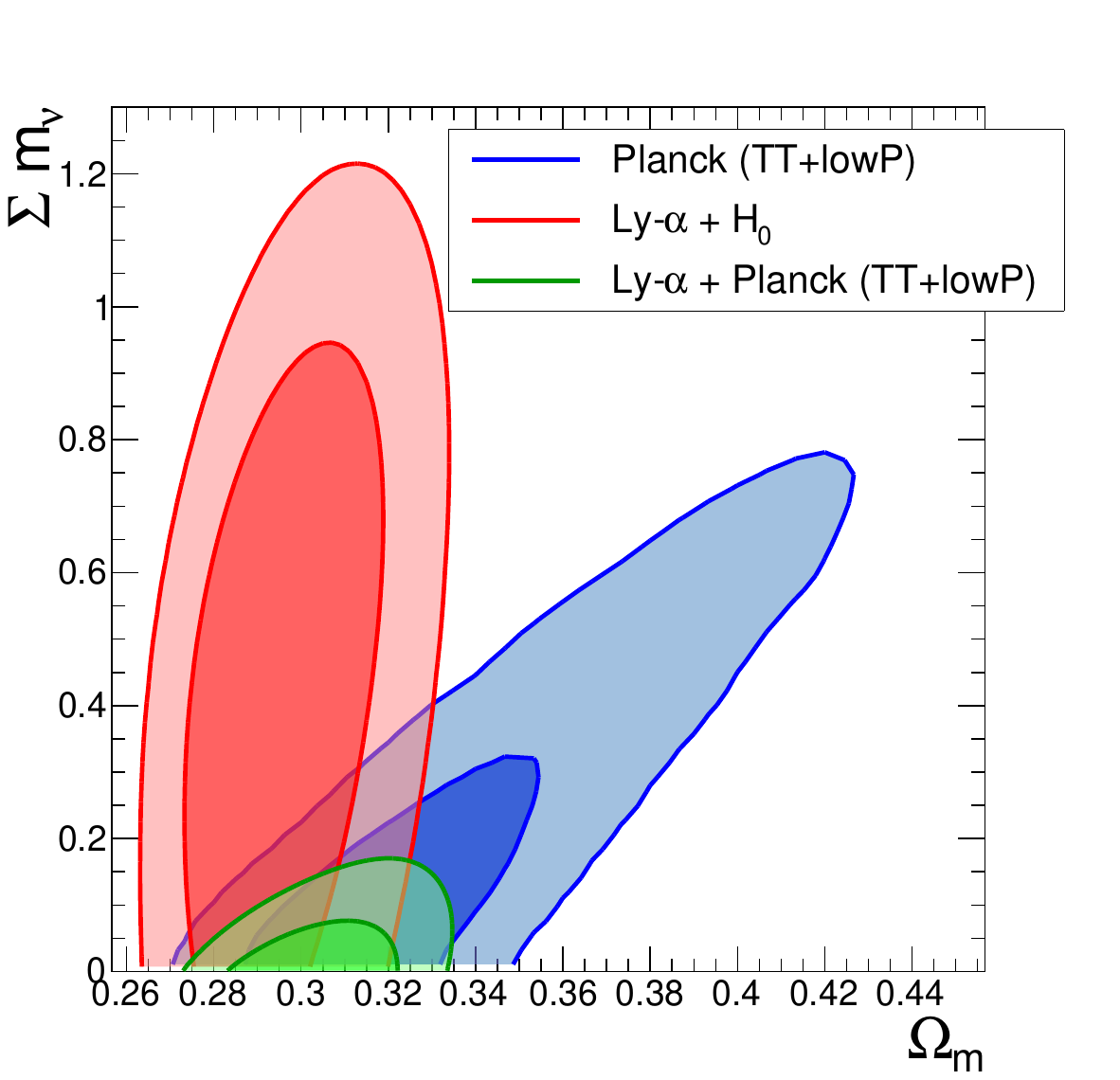,width = 5.0cm}
\epsfig{figure= 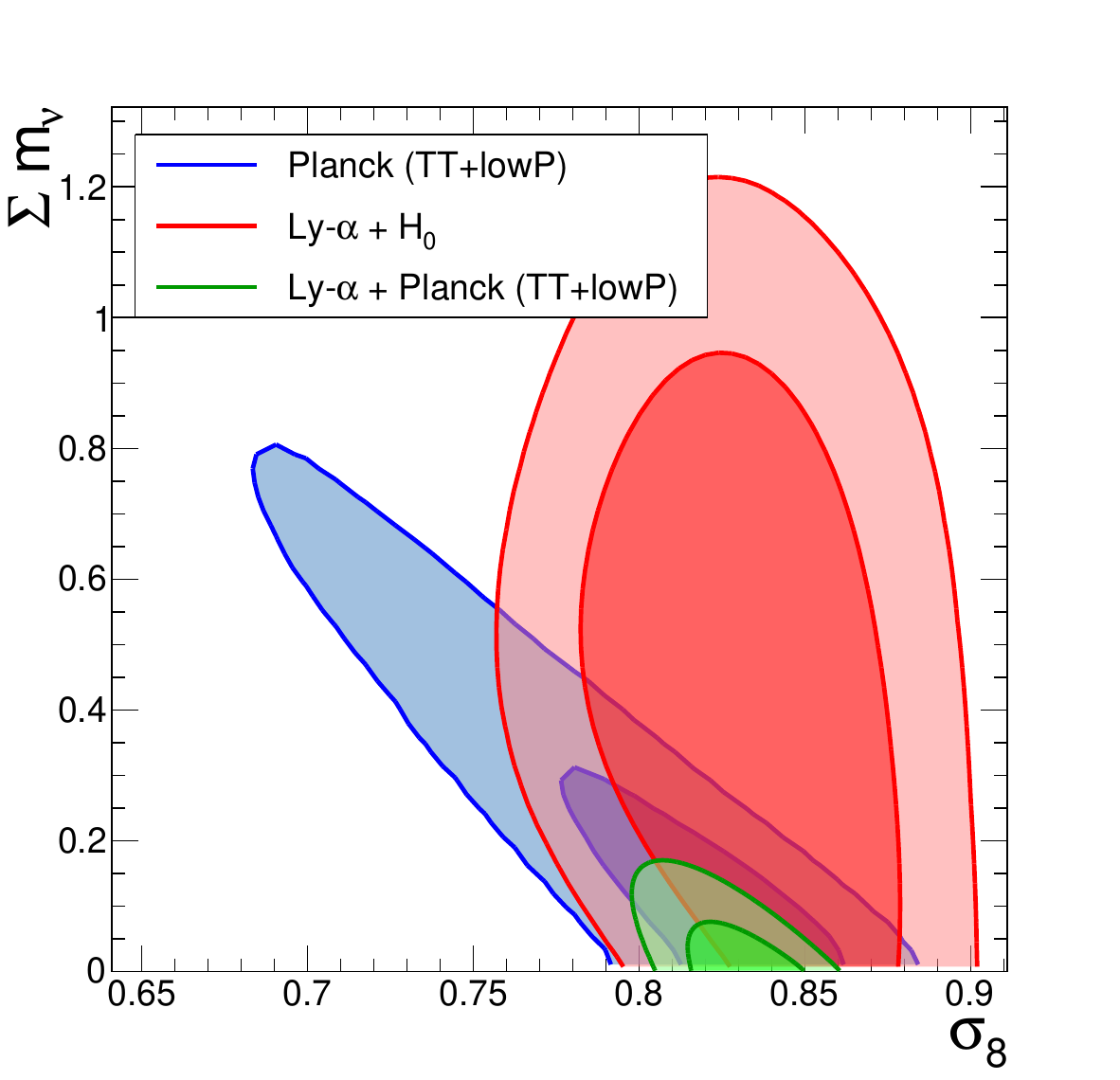,width = 5.0cm}
\caption{\it 2D confidence level contours for  the $(\sigma_8 , n_s)$ ,  $(\Omega_m , \sum m_\nu)$ and $(\sigma_8 , \sum m_\nu)$   cosmological parameters.  The 68\% and 95\% confidence contours are obtained with different combinations of the BOSS Ly$\alpha$ data presented in section~\ref{sec:LyaAlone}  of the Gaussian constraint $H_0 = 67.4 \pm 1.4~{\rm km~s^{-1}~Mpc^{-1}}$ and of Planck 2015 data (TT+lowP). }
\label{fig:OmegamMnuLyaPlanck}
\end{center}
\end{figure}

\begin{figure}[htbp]
\begin{center}
\epsfig{figure= 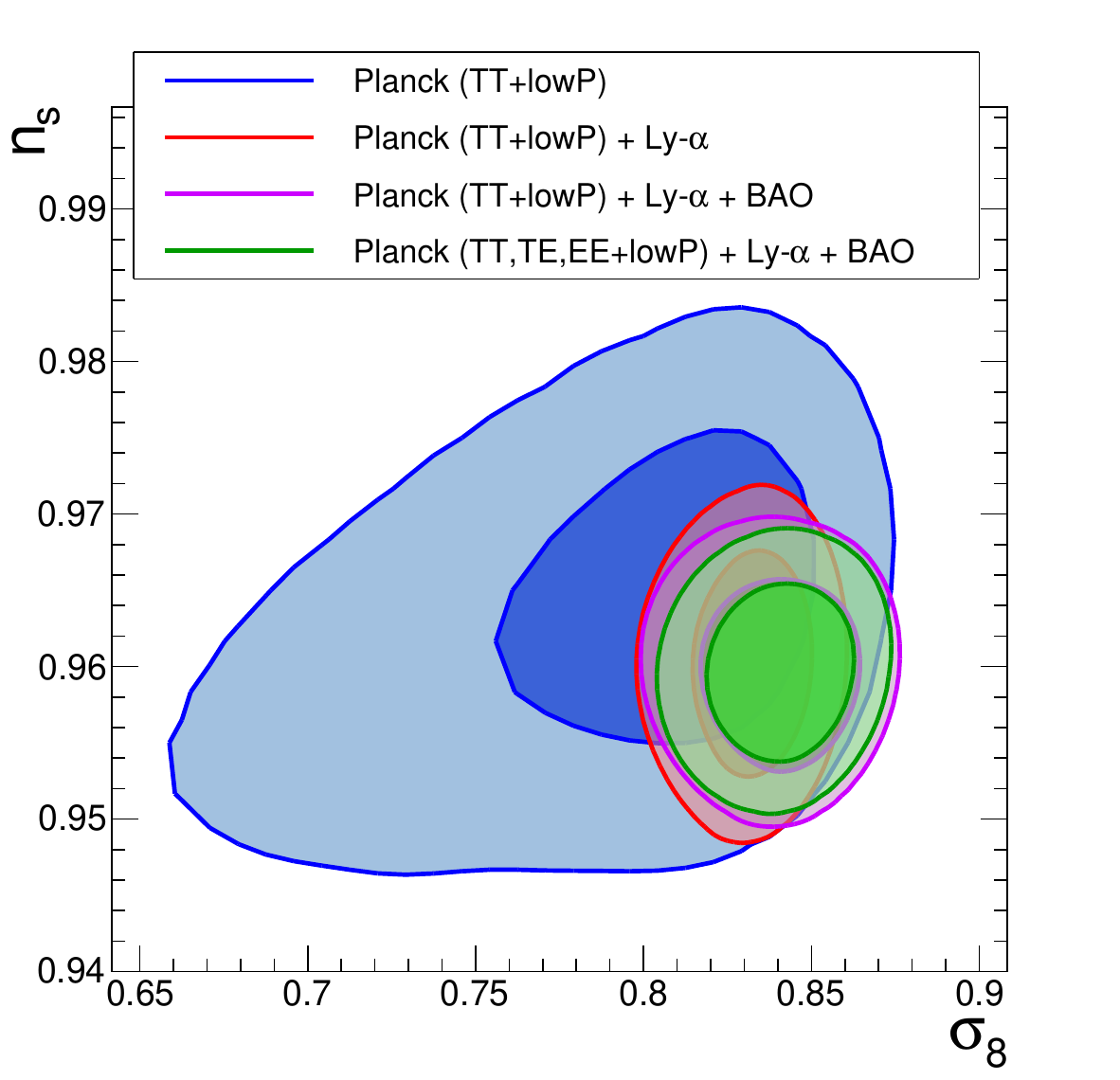,width = 5.0cm}
\epsfig{figure= 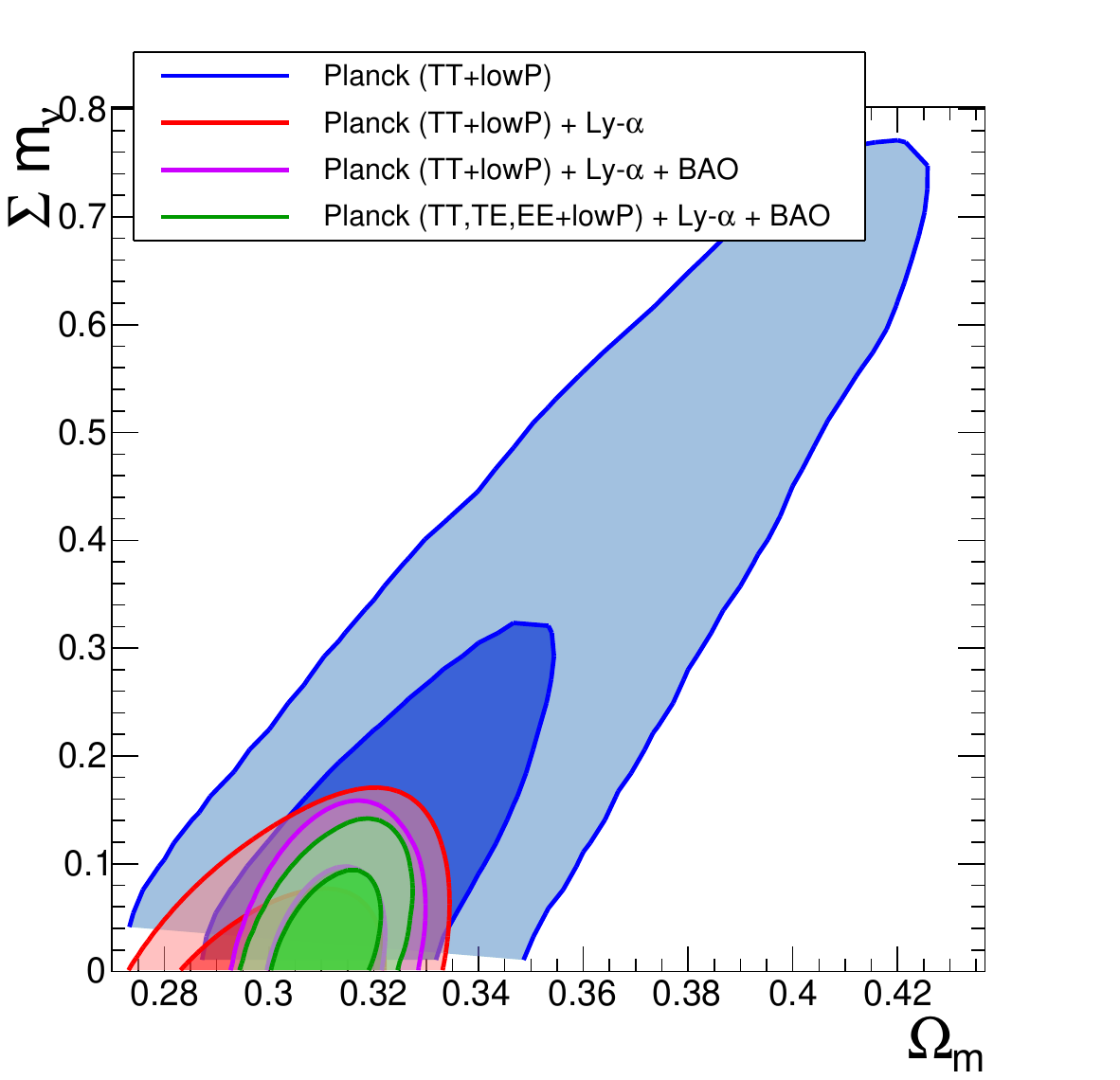,width = 5.0cm}
\epsfig{figure= 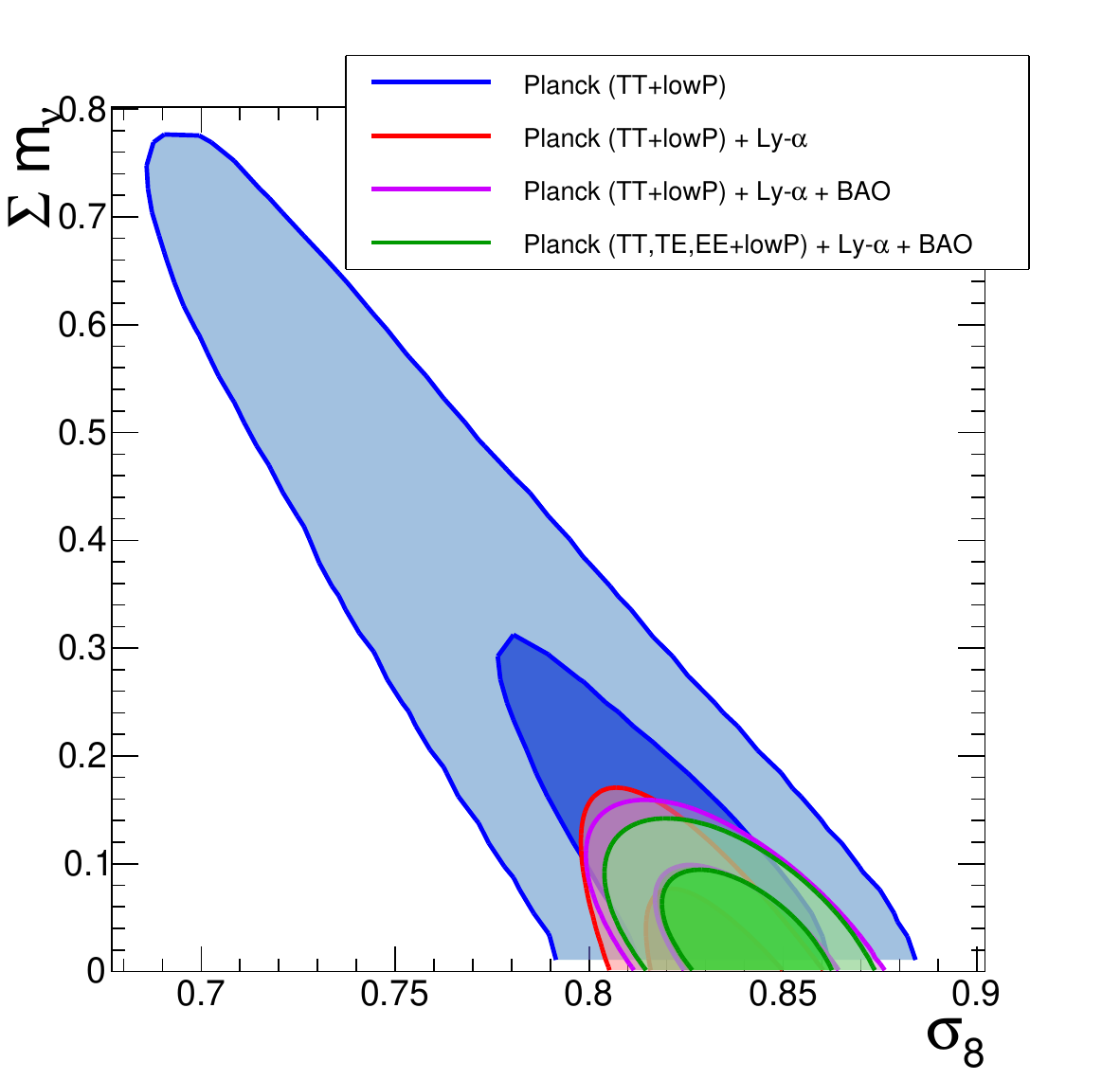,width = 5.0cm}
\caption{\it 2D confidence level contours for  the $(\sigma_8 , n_s)$ , $(\Omega_m , \sum m_\nu)$ and $(\sigma_8 , \sum m_\nu)$  cosmological parameters.  The 68\% and 95\% confidence contours are obtained for four combinations, starting with Planck 2015 data  (TT+lowP) alone,  then  adding  the BOSS Ly$\alpha$ data,  the BAO data and finally the high-$\ell$ polarization spectra for Planck (TE and EE).  }
\label{fig:OmegamMnuVariousPlanck}
\end{center}
\end{figure}

In this section, we  combine the  Ly$\alpha$ likelihood (imposing no constraint on $H_0$) with the likelihood of Planck 2015 data that we derive from the central values and covariance matrices available in the official 2015 Planck repository. As in the previous section, we  focus on the base $\Lambda$CDM$\nu$ model, and we derive constraints on $\sigma_8$, $n_s$, $\Omega_m$, $H_0$ and $\sum m_\nu$. Results are shown in table.~\ref{tab:fit_Lya_CMB}.  Column (1) is the same as column (2) of table~\ref{tab:fit_Lya} and recalls the results for Ly$\alpha$ alone. Column (2) is for the combined set of Ly$\alpha$ and the base configuration we chose for Planck data, i.e. TT+lowP (cf. details in Sec.~\ref{sec:cmb}). The last two columns (columns 3--4)  include BAO data in addition, and in column (4) we extend the CMB measurements to TT+TE+EE+lowP.
 We illustrate  the main 2D contours on cosmological parameters in figures~\ref{fig:OmegamMnuLyaPlanck} and \ref{fig:OmegamMnuVariousPlanck}. 
 
\begin{table}[htbp]

\caption{\it Best-fit value and 68\% confidence levels of the cosmological parameters of the model fitted to the flux power spectrum $P(k_i,z_j)$ measured with the BOSS Ly$\alpha$ data combined with several other data sets.}

\begin{center}

\begin{tabular}{lcccc}
\hline
 &  (1) Ly$\alpha$   & (2)  Ly$\alpha$ 
 & (3)  Ly$\alpha$ & (4) Ly$\alpha$  \\
 
Parameter &  + $H_{0}^\mathrm{Gaussian}$  &  + Planck {\scriptsize TT+lowP}
&   + Planck {\scriptsize TT+lowP } &  + Planck {\scriptsize TT+TE+EE+lowP} \\
 
 & {\scriptsize ($H_0 = 67.3 \pm 1.0$)}  &  & + BAO & + BAO   \\
 \hline \\[-10pt]
 
$\sigma_8$ &  $0.831\pm0.031$ & $0.833 \pm 0.011$ & $0.845\pm 0.010$ & $ 0.842\pm 0.014$\\[2pt]
$n_s$ &  $0.938\pm0.010$  &  $0.960\pm0.005$ &  $0.959 \pm 0.004$ & $0.960 \pm 0.004$\\[2pt]
$\Omega_m$  &  $0.293\pm0.014$ &  $0.302\pm0.014 $ &  $0.311\pm0.014$& $0.311 \pm 0.007$\\[2pt]
$H_0$~{\scriptsize(${\rm km~s^{-1}~Mpc^{-1}}$)}   & $67.3\pm1.0$ &   $68.1 \pm 0.9$ & $67.7\pm 1.1$ & $67.7 \pm 0.6$ \\[2pt]
$\sum \! m_\nu$~{\scriptsize(eV)} &$<1.1 $~{\scriptsize(95\% CL)} & $<0.12$ ~{\scriptsize(95\% CL)}& $< 0.13$~{\scriptsize(95\% CL)} & $<0.12$~{\scriptsize(95\% CL)} \\[2pt]
\hline\\[-11pt]
Reduced $\chi^2$&0.99 & 1.04& 1.05& 1.05\\[2pt]
\hline
\end{tabular}
\end{center}
\label{tab:fit_Lya_CMB}
\end{table}

The main point to note is the  excellent agreement between the results derived from the combination of Ly$\alpha$ data with   different sets of CMB and BAO data  (columns 2--4). The consistency between these additional data  was already discussed in the Planck cosmology paper~\cite{Planck2015}. The agreement  with the values  obtained from Ly$\alpha$ data alone is also remarkable, and significantly improved compared to the  previous Ly$\alpha$ analysis (Paper~I): the downward shift of $\sigma_8$ and upward shift of $n_s$ noted in the previous section bring these two parameters  closer to the value measured from CMB data. The upper bound on total  neutrino mass is tighter than in  Paper~I. We now constrain $\sum m_\nu$ to be less than 0.12~eV (at 95\% C.L.) from Ly$\alpha$ and Planck TT+lowP, instead of 0.15~eV in Paper~I. 
The  improvement  of our limit is largely dominated by our improved calibration of the hydrodynamical simulations, and in particular by our new  model of  the   bias induced by the  splicing technique.  The use of two additional redshift bins, and the update to Planck 2015 data also contribute to this improvement, although to a lesser extent. The main restriction to further improvement comes from the additional flexibility we allowed in the  parameters describing the temperature evolution of the IGM. 
Our new constraint is  much closer to the inverted-hierarchy lower bound of 0.10~eV than current CMB-based limits. For comparison, Planck (TT+lowP) alone constrains the sum of the neutrino masses to $\sum m_\nu<0.72~eV$,  Planck (TT+lowP) + BAO to $\sum m_\nu<0.21$~eV, and Planck (TT+TE+EE+lowP) + BAO to $\sum m_\nu<0.17$~eV, all at 95\% CL.

The small  tension on $n_s$ that was discussed in Paper~I is still present, at the 2.3~$\sigma$ level, although no longer affected by the presence or not of additional BAO data. We will study in Sec.~\ref{sec:primordial} possible extensions to the base $\Lambda$CDM$\nu$ models to incorporate this  tension, and evaluate its impact  on inflation models and a possible running of the scalar spectral index. The tension on $n_s$ has little effect on the constraint on $\sum m_\nu$ because  of the mild  correlation between these two parameters  (48\% in Ly$\alpha$, -45\% in Planck TT+lowP).  As in Paper~I, $\sum m_\nu$ is mostly correlated to $\sigma_8$ (-48\% in Ly$\alpha$, -95\% in Planck TT+lowP), and to $\Omega_m$ (52\% in Ly$\alpha$, 92\% in Planck TT+lowP).    
  
It should be noted that the combination of CMB and Ly$\alpha$
is a very efficient way of constraining cosmological parameters,
especially $\sum m_\nu$.  As one can see in Figure 5, the Planck
and \lya+$H_0$ contours in the $\sum m_\nu-\Omega_m$ and
$\sum m_\nu-\sigma_8$ planes are complementary.
The Ly$\alpha$ data constrain $\Omega_m$ and $\sigma_8$ largely
independently of $\sum m_\nu$ because they have different
impact on the shape of the power spectrum (see discussion in
\S 5.1 of Paper~I).
For the Planck constraints, high $\sum m_\nu$ corresponds
to low $\sigma_8$ because of the suppression of power on
small scales by neutrino free streaming.  The positive
correlation between $\Omega_m$ and $\sum m_\nu$ is more
subtle: with $\Omega_c h^2$ and $\Omega_b h^2$ well constrained
by the acoustic peaks, raising $\sum m_\nu$ increases the matter
density at low redshift after neutrinos become non-relativistic,
and within $\Lambda$CDM this requires a decrease in $h$ to maintain the well determined angular diameter distance to last scattering,
and this in turn corresponds to higher $\Omega_m$
(see, e.g., \S 6.4 of \cite{Planck2015}).
The end result is that the \lya\ and Planck contours
intersect only near $\sum m_\nu=0$.
As shown in Figure~\ref{fig:OmegamMnuVariousPlanck}, adding polarization or BAO to the
Planck + Ly$\alpha$ contours does not lead to significant further
improvement of the constraints in these planes, at least
within the $\Lambda$CDM framework.


\subsection{Reionization optical depth  from Ly$\alpha$ data and other probes}
\label{sec:tau}
In this section, we use the  same  Planck data sets as in Sec.~\ref{sec:LyaCMB}, and we now consider constraints that can be set on the optical depth  $\tau$.  More precisely, we study the improvement, over what CMB alone can do, provided by the addition of Ly$\alpha$ and BAO data sets, through correlations  and tightened constraints on the cosmological parameters each is sensitive to. 

\begin{figure}[htbp]
\begin{center}
 \epsfig{figure= 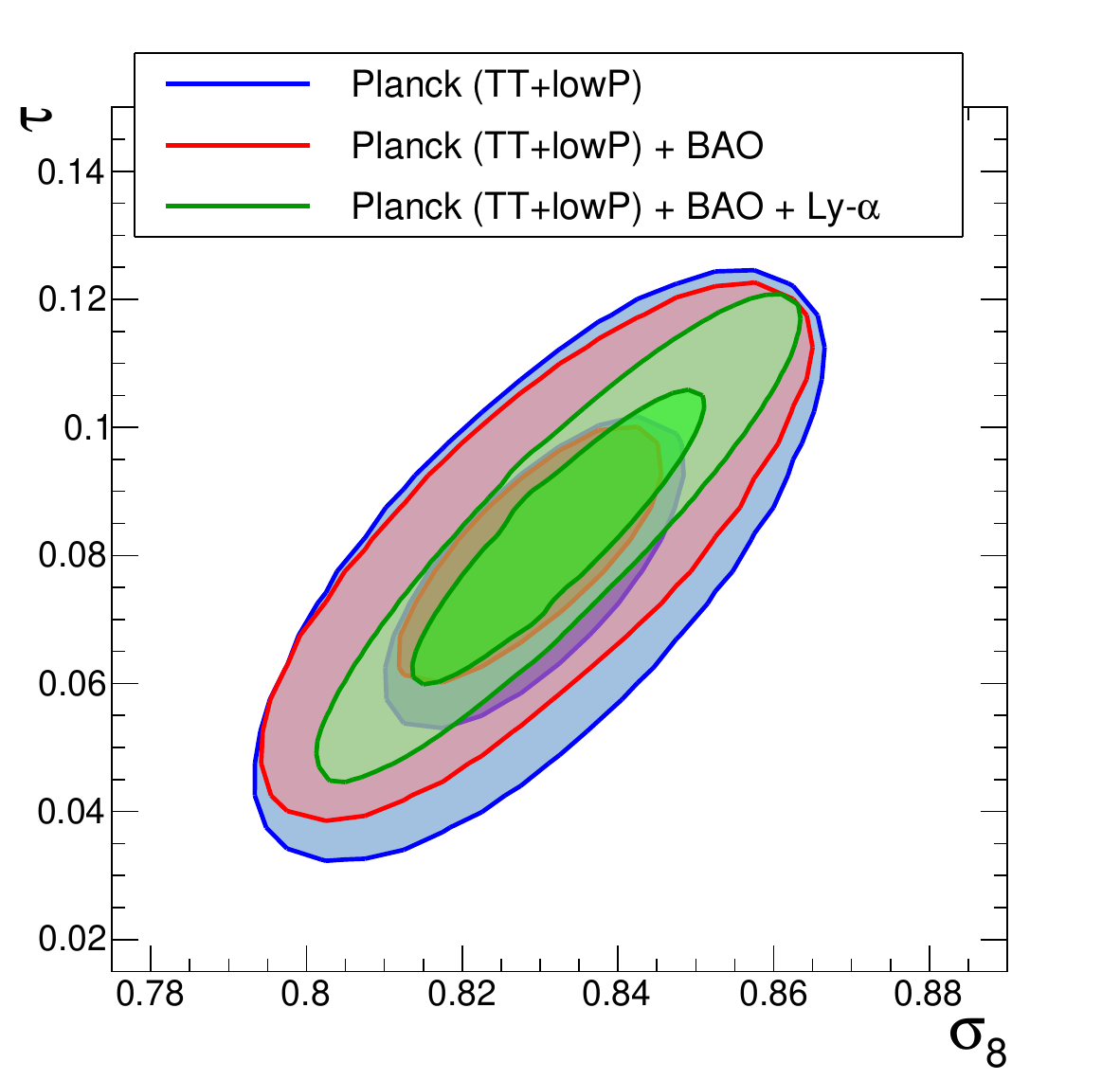 ,width = 7.5cm}
\epsfig{figure= 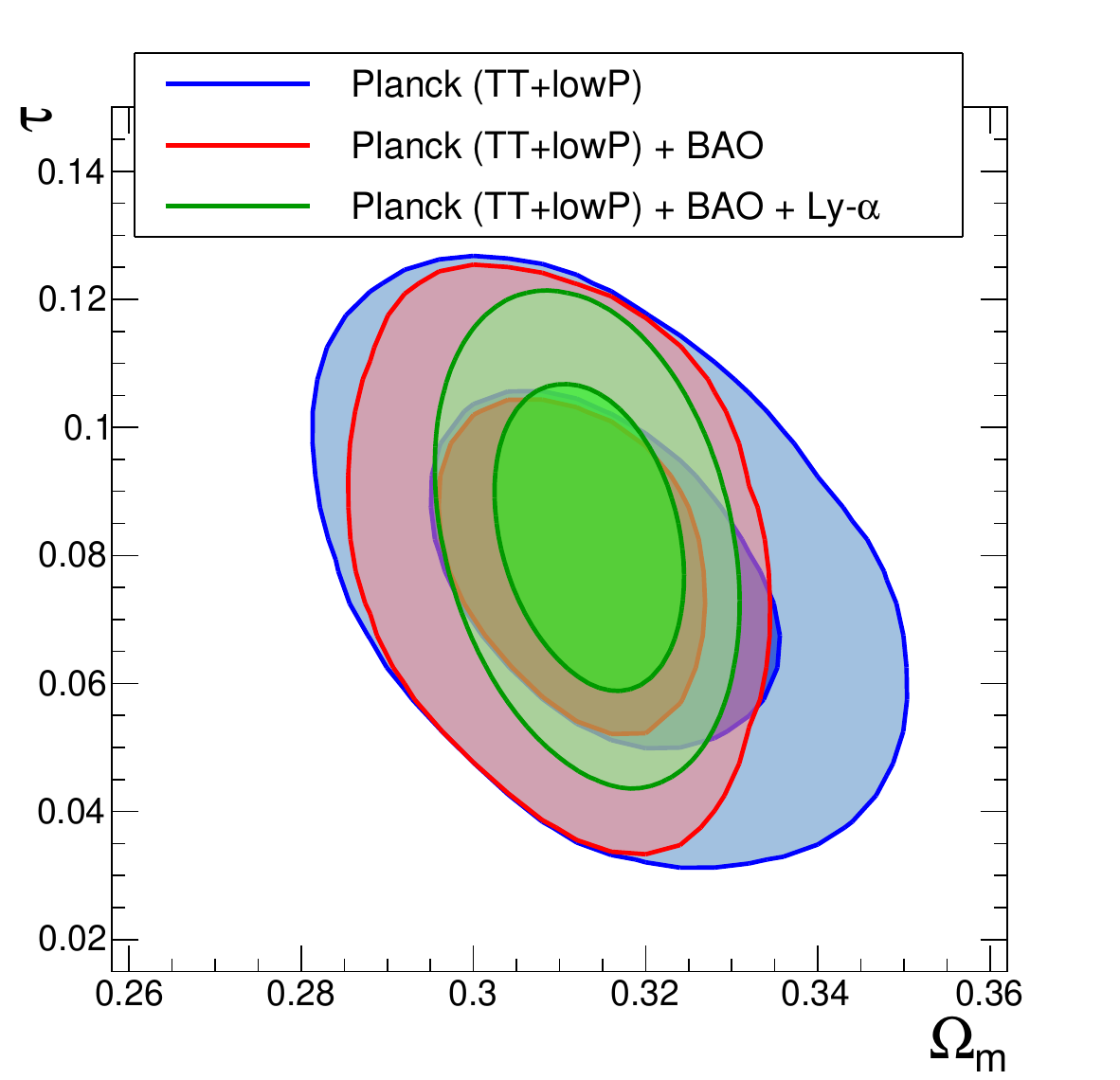, width = 7.5cm}
\caption{\it  2D confidence level contours for  the $(\sigma_8 , \tau)$ and $(\Omega_m , \tau)$ .  The 68\% and 95\% confidence contours are obtained for three combinations, starting with Planck 2015 data  (TT+lowP) alone,  then by adding  the BAO data and finally  the BOSS Ly$\alpha$ data. }
\label{fig:tauPlanck}
\end{center}
\end{figure}

The optical depth is mainly constrained by the CMB through two effects~\cite{PlanckCollaboration2013,Planck2015}.
First, small-scale fluctuations in the CMB are damped by Thomson scattering from free electrons produced at reionization. This scattering suppresses the amplitude of the CMB acoustic peaks by $e^{-2\tau}$. Therefore Planck measures in the TT power spectrum the damped amplitude $e^{-2\tau}A_s$ on small scales. This effect introduces a strong correlation between $\tau$ and $\sigma_8$. 
Secondly, $\tau$ can be measured from CMB polarization in the multipole range $\ell =2-30$. So far the 	analysis presented by the Planck collaboration~\cite{Planck2015}   used the low-resolution LFI 70GHz maps for small $\ell$.  The $\tau$ measurement is improved by the addition of the polarization information, but this improvement does not break the degeneracy between $\tau$ and $\sigma_8$, as shown on  figure~\ref{fig:tauPlanck}.  We can expect further improvement once the measurements of  low multipoles with HFI polarization are released. 

We give below the constraints on the optical depth $\tau$ and on the redshift at  reionization $z_{\rm re}$ in the framework of the concordance model $\Lambda$CDM for four combinations of data sets. The Planck~(TT+lowP) and the Planck  (TT+lowP) + BAO cases are taken from \cite{Planck2015}; we add the Ly$\alpha$ likelihood to either case to produce the last two configurations.
\begin{eqnarray*}
\tau =  0.078 \pm  0.019 \hspace{1cm} &  z_{\rm re} =  9.9 ^{+1.8}_{-1.6} \hspace{1cm} & {\rm Planck~(TT+lowP) }\\
\tau =  0.080 \pm 0.016\hspace{1cm} &  z_{\rm re} =  10.2 ^{+1.6}_{-1.5} \hspace{1cm} & {\rm Planck~(TT+lowP) + Ly}\alpha \\
\tau =  0.080 \pm 0.017\hspace{1cm} &  z_{\rm re} =  10.1 ^{+1.6}_{-1.5} \hspace{1cm} & {\rm Planck~(TT+lowP) + BAO}\\
\tau =  0.083 \pm 0.015\hspace{1cm} &  z_{\rm re} =  10.4 ^{+1.5}_{-1.4} \hspace{1cm} & {\rm Planck~(TT+lowP) + BAO + Ly}\alpha \\
\end{eqnarray*}
The  uncertainties on $\tau$  and $z_{\rm re}$ are tightened more significantly by adding Ly$\alpha$ than by including BAO. The values of $z_{\rm re}$ are slightly higher than for Planck alone when  combining either with BAO or/and Ly$\alpha$. The gain in precision is clearly due to the correlation between the parameters $\tau$ and $\sigma_8$  and to a lesser extent to the correlation between $\tau$ and $\Omega_m$  (see figure~\ref{fig:tauPlanck}).

\subsection{Primordial fluctuations from Ly$\alpha$ data and other probes}
\label{sec:primordial}
In this section, we use the same  data sets as in Sec.~\ref{sec:LyaCMB} but we extend the model to allow for additional parameters. In particular, the  small tension between the values of $n_s$ preferred by Ly$\alpha$ or Planck data motivates a combined fit allowing $n_s$ to vary with scale. We thus introduce ${\mathrm d}n_s/{\mathrm d}\ln k$ by using the corresponding Planck chains and adapting the Ly$\alpha$ likelihood to include a running of $n_s$. 
We choose a pivot scale $k_0=0.05~{\rm Mpc}^{-1}$, the same as in the analyses led by the Planck  collaboration in order to allow  direct comparisons. The scalar mode power spectrum is then parameterized by a power law with
\begin{equation}
P_s=\biggl(\frac{k} {k_0}\biggr)^{n_s-1+\frac{1} {2} {\mathrm d}n_s/{\mathrm d}\ln k \ln(k/k_0)} \; .
\label{eq:running}
\end{equation}
As ${\mathrm d}n_s/{\mathrm d}\ln k$ is not included in the grid, we take into account the variation with scale  of $n_{s }$ in the Ly$\alpha$ data in the following way: we replace  $n_s$ in the Ly$\alpha$ likelihood by $n_{s, \, Ly\alpha }(k) = n_s +1/2 \cdot{\mathrm d}n_s/{\mathrm d}\ln k \cdot \ln(k/k_0)$. With this approximation, we can fit CMB and Ly$\alpha$ data with common $n_s$ and 
${\mathrm d}n_s/{\mathrm d}\ln k$ parameters. The pivot scale $k_0$ is approximately in the middle of the logarithmic range of the scales probed by Planck. The Ly$\alpha$ forests cover scales ranging from $\sim 0.07~{\rm Mpc}^{-1}$ to $\sim 1.7~{\rm Mpc}^{-1}$, with a pivot near $k_{Ly\alpha} \sim  0.7~{\rm Mpc}^{-1}$. The constraint that we can derive on the running index ${\mathrm d}n_s/{\mathrm d}\ln k$ comes mostly from the different levels of the power spectra at the CMB and Ly$\alpha$ pivot scales $k_0$ and $k_{Ly\alpha}$.

On the theoretical side, the simplest inflationary models predict that the running of the spectral index should be of second order in inflationary slow-roll parameters and therefore small, $|{\mathrm d}n_s/{\mathrm d}\ln k|\sim (n_s-1)^2\sim 10^{-3}$~\cite{Kosowsky1995}. Nevertheless, it is possible to accommodate a larger scale dependence of $n_s$,  by adjusting the third derivative in the inflaton potential, for instance. A negative running of order $10^{-2}$ can also be implemented in inflation models with oscillations in the inflaton potential, using axion monodromy~\cite{Minor2015}.
On the experimental side, recent CMB experiments have a mixed history of  null-results and a-few-sigma detections of running of the scalar index. The final 9-year WMAP analysis found no evidence of running using WMAP alone, with ${\mathrm d}n_s/{\mathrm d}\ln k = -0.019\pm 0.025$ at 68\% CL, while the combination of WMAP data with the first data releases from ACT and SPT found a negative running at nearly the $2\sigma$ level with ${\mathrm d}n_s/{\mathrm d}\ln k = -0.022\pm 0.012$ \cite{Hinshaw2013}. The ACT 3-year release measured ${\mathrm d}n_s/{\mathrm d}\ln k = -0.003\pm 0.013$ when combining with WMAP-7~\cite{Sievers2013}. A negative running was detected at just over $2\sigma$ by SPT, ${\mathrm d}n_s/{\mathrm d}\ln k = -0.024\pm 0.011$~\cite{Hou2014}. The Planck 2015 results, while   roughly consistent with zero running of the scalar spectral index,  indicate a $\sim 1\sigma$ preference for negative running, ${\mathrm d}n_s/{\mathrm d}\ln k = -0.0084\pm 0.0082$, with a slightly lower significance when adding HFI polarization or BAO (${\mathrm d}n_s/{\mathrm d}\ln k = -0.0057 \pm 0.0070$), and on the contrary  a slightly higher significance when tensor fluctuations are also allowed in addition to running (${\mathrm d}n_s/{\mathrm d}\ln k = -0.0126\pm 0.0090$).

\begin{figure}[htbp]
\begin{center}
\epsfig{figure= 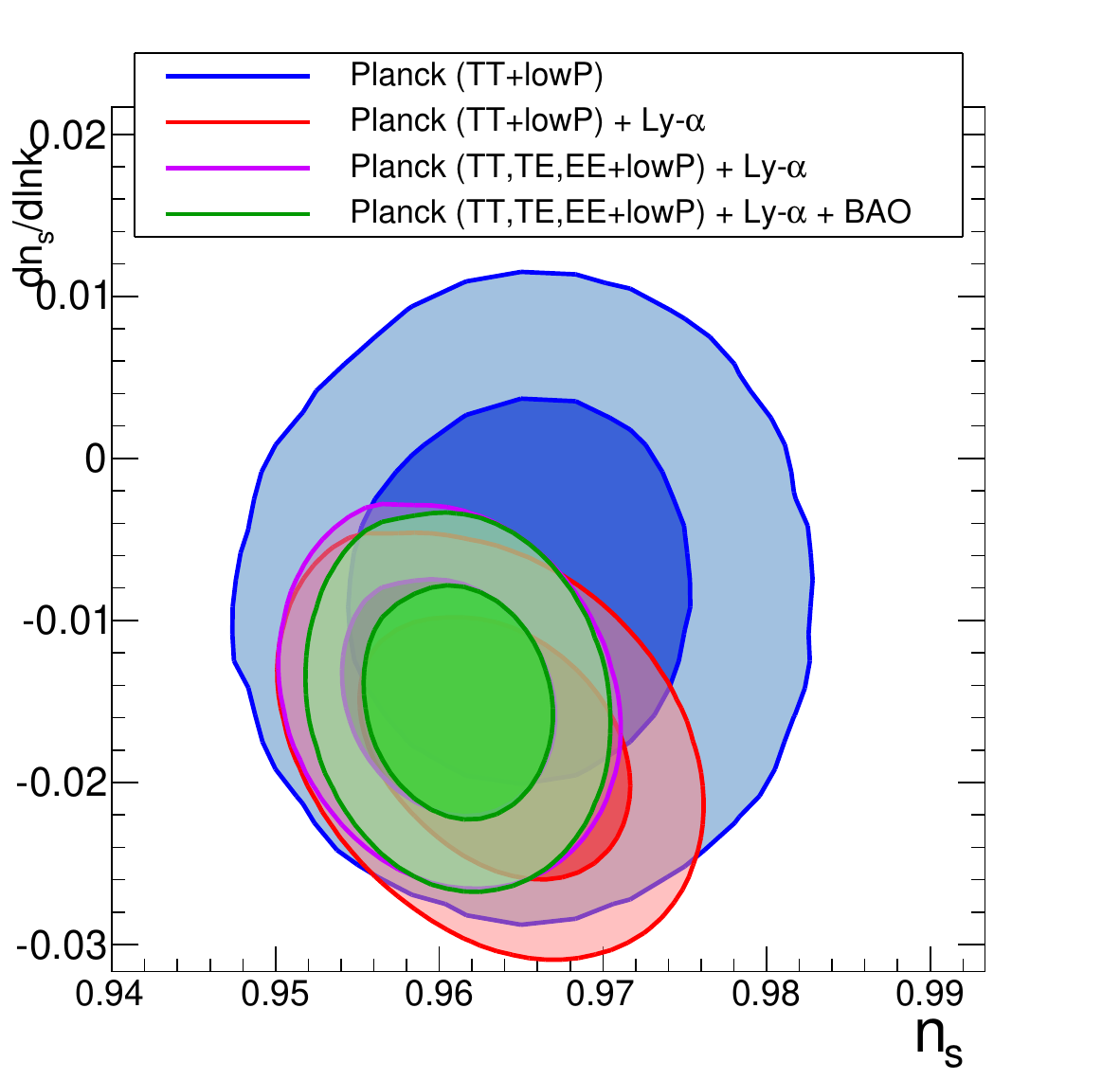,width = 5.0cm}
\epsfig{figure= 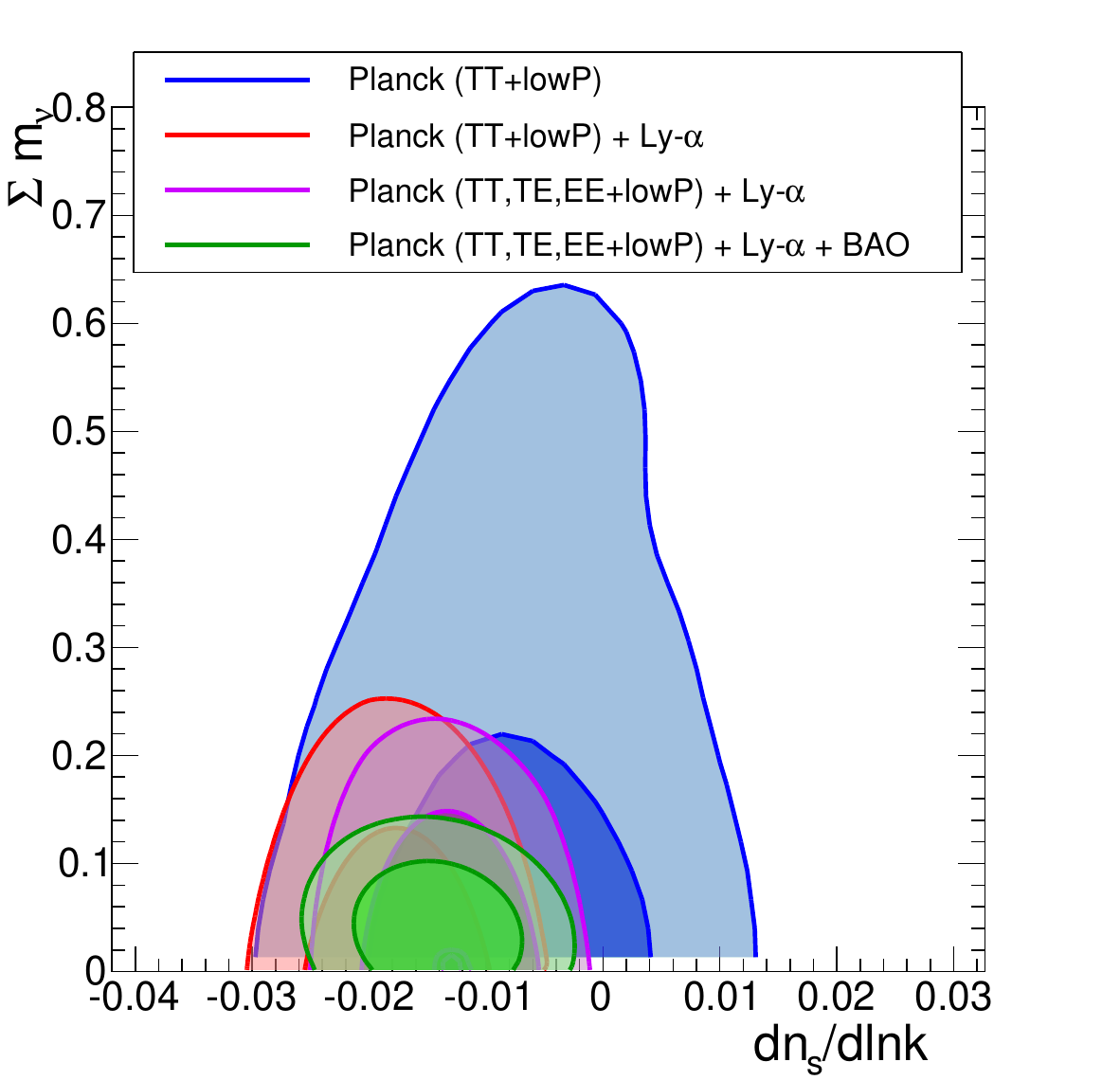,width = 5.0cm}
\epsfig{figure= 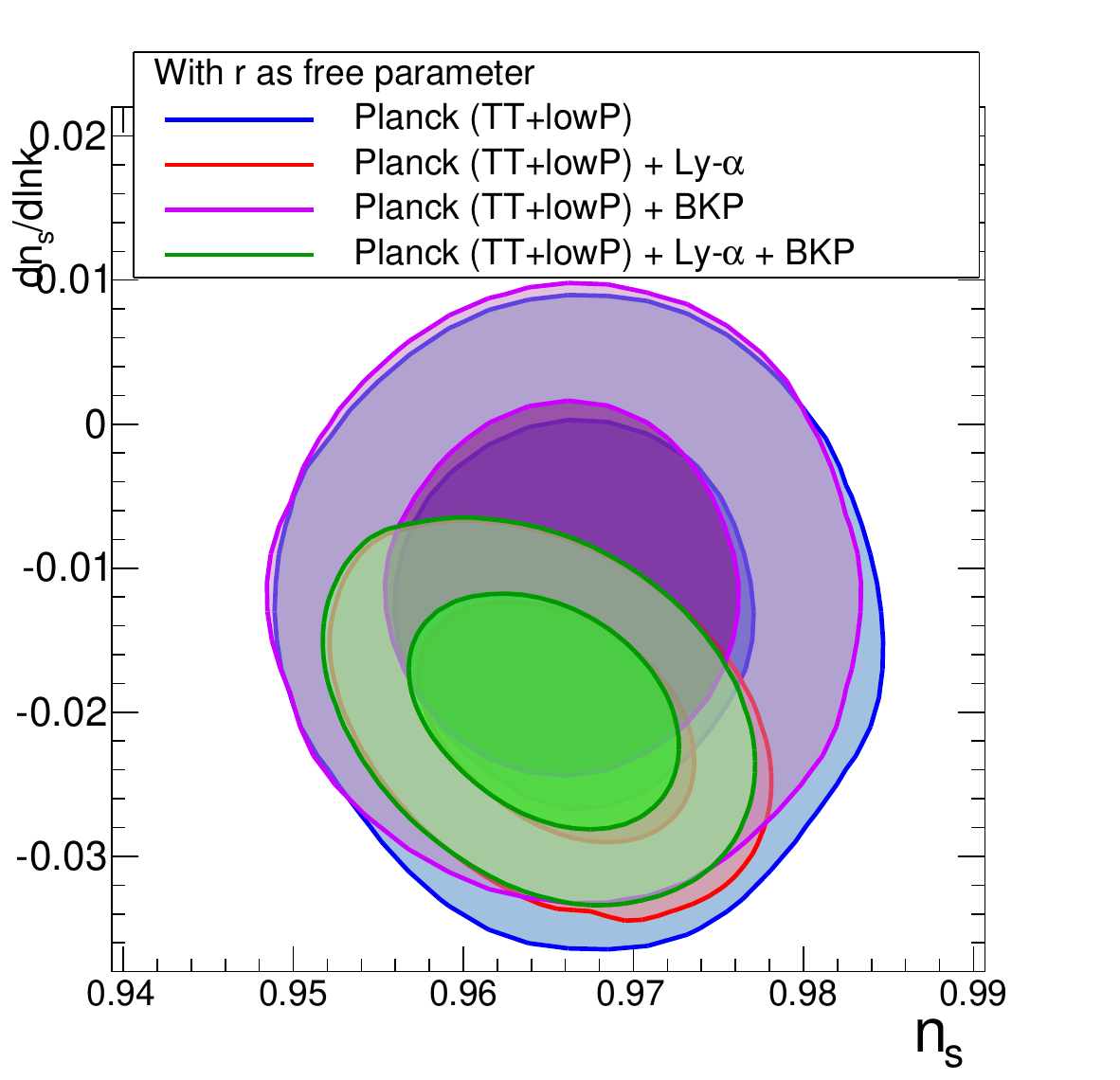,width = 5.0cm}3
\caption{\it Constraints on the scalar spectral index $n_s$, the running ${\mathrm d}n_s/{\mathrm d}\ln k$, and  $\sum m_\nu$. Left and middle:   68\% and 95\% confidence contours  obtained for four combinations --  Planck 2015  TT+lowP data alone,  then  adding   BOSS Ly$\alpha$,   high-$\ell$ polarization  from Planck (TE and EE) and finally  BAO data. Left plot is for $\Lambda$CDM with running,  middle plot for $\Lambda$CDM$\nu$ with running. Right: the  tensor-to-scalar ratio $r$ is floated. The 68\% and 95\% confidence contours are obtained for four combinations -- Planck 2015 TT+lowP,   Planck 2015 TT+lowP and BOSS Ly$\alpha$,  Planck 2015 TT+lowP and  BICEP2/Keck Array  and finally all the above together. }
\label{fig:running}
\end{center}
\end{figure}

In the table below, we give the values of $n_s$ and of ${\mathrm d}n_s/{\mathrm d}\ln k$ that we find for the $\Lambda$CDM model with a tensor-to-scalar ratio $r=0$. The corresponding 2D contours are illustrated in figure~\ref{fig:running}, left plot.
\begin{eqnarray*}
n_s = 0.965\pm0.007 \hspace{.6cm} &  {\mathrm d}n_s/{\mathrm d}\ln k =  -0.0084_{-0.0081}^{+0.0082} \hspace{.6cm} & {\rm Planck~(TT+lowP) }  
\nonumber \\
n_s =0.964 \pm 0.005\hspace{.6cm} &  {\mathrm d}n_s/{\mathrm d}\ln k =  -0.0178_{-0.0048}^{+0.0054} \hspace{.6cm}& {\rm Planck~(TT+lowP) + Ly}\alpha 
\nonumber \\
n_s = 0.960\pm 0.004\hspace{.6cm} &  {\mathrm d}n_s/{\mathrm d}\ln k =  -0.0149_{-0.0048}^{+0.0050} \hspace{.6cm}& {\rm Planck~(TT,TE,EE+lowP) + Ly}\alpha 
\nonumber \\
n_s = 0.961\pm 0.004\hspace{.6cm} &  {\mathrm d}n_s/{\mathrm d}\ln k =  -0.0152 _{-0.0045}^{+0.0050}\hspace{.6cm} & {\rm Planck~(TT,TE,EE+lowP) +BAO + Ly}\alpha 
\end{eqnarray*}

Allowing a running of $n_s$ improves  the fit $\chi^2$ by $\sim 10$  compared to the results obtained for the same set of data but without running. This is driven by the fact that a negative running of order $10^{-2}$ is favored both by  Planck data alone, and by the tension on $n_s$ between Planck and Ly$\alpha$ data sets.  The four different combinations of Planck, Ly$\alpha$ and BAO data shown above all indicate a preference for a negative running at the $\sim 3\sigma$ level. When allowing for running, the best-fit value of $n_s$ is  in excellent agreement with the value favored by Planck data alone. 
The best-fit value of $ {\mathrm d}n_s/{\mathrm d}\ln k$ is also in agreement (at about $1~\sigma$) with the value measured from CMB data alone by the Planck collaboration, although the origin of this detection is to first order only dependent on the different values of $n_s$ in Ly$\alpha$ and CMB data, and not on the value of running measured from either probe alone (see discussion in Sec.~\ref{sec:discussion}). 
The  detection of running in the combined fit, however, could be the result of a coincidence between a $\sim 1~\sigma$ effect in Planck, due to the mismatch between the high and low multipoles in the temperature power spectrum on the one hand, and a $\sim 2.3~\sigma$ effect in Ly$\alpha$ data, possibly coming from an unidentified systematic bias on the other hand.  

As  the global $\chi^2$ is clearly improved by letting  ${\mathrm d}n_s/{\mathrm d}\ln k$ free, it is interesting to study the impact of this extra parameter on the determination of  $\sum m_\nu$  in the base $\Lambda$CDM$\nu$ model with running. As shown on the middle plot of the figure~\ref{fig:running}, the correlation between ${\mathrm d}n_s/{\mathrm d}\ln k$ and $\sum m_\nu$ is small. In the table below, we give the values of $\sum m_\nu$ and of ${\mathrm d}n_s/{\mathrm d}\ln k$ for the four configurations already studied. 
\begin{eqnarray*}
 \Sigma m_\nu<0.65\,{\rm eV}\,(95\%{\rm CL})  \hspace{.6cm} &  {\mathrm d}n_s/{\mathrm d}\ln k =  -0.0078_{-0.0083}^{+0.0084} \hspace{.6cm} & {\rm Planck~(TT+lowP) }  \\
 \Sigma m_\nu<0.19\,{\rm eV}\,(95\%{\rm CL})  \hspace{.6cm} &  {\mathrm d}n_s/{\mathrm d}\ln k =  -0.0178_{-0.0052}^{+0.0054} \hspace{.6cm}& {\rm Planck~(TT+lowP) + Ly}\alpha \\
 \Sigma m_\nu<0.19\,{\rm eV}\,(95\%{\rm CL})  \hspace{.6cm} &  {\mathrm d}n_s/{\mathrm d}\ln k =  -0.00135_{-0.0050}^{+0.0046} \hspace{.6cm}& {\rm Planck~(TT,TE,EE+lowP) + Ly}\alpha \\
 \Sigma m_\nu<0.12\,{\rm eV}\,(95\%{\rm CL})  \hspace{.6cm} &  {\mathrm d}n_s/{\mathrm d}\ln k =  -0.00141 _{-0.0048}^{+0.0047}\hspace{.6cm} & {\rm Planck~(TT,TE,EE+lowP) +BAO + Ly}\alpha 
\end{eqnarray*}
We obtain an impressive improvement on the bound on $\sum m_\nu$ by including Ly$\alpha$ on top of CMB data. However, the gain is slightly lower than when ${\mathrm d}n_s/{\mathrm d}\ln k$ is fixed to zero. It may be an indication that part of the improvement obtained with Ly$\alpha$ data is due to the small tension on the value of $n_s$   between  CMB and Ly$\alpha$. Finally, even with ${\mathrm d}n_s/{\mathrm d}\ln k$ as an additional free parameter, we obtain the same constraint $\sum m_\nu<0.12$~eV (95\% C.L.)  when combining  the three probes CMB, Ly$\alpha$ and BAO.

Letting $r$ free does not change qualitatively the constraint on ${\mathrm d}n_s/{\mathrm d}\ln k$, as shown in the table below and as illustrated in the right plot of   figure~\ref{fig:running}. In the combination, we also  introduce the BICEP2/Keck Array-Planck (BKP) data set for reasons explained hereafter.
\begin{eqnarray*}
n_s = 0.967\pm0.007 \hspace{.6cm} &  {\mathrm d}n_s/{\mathrm d}\ln k =  -0.0126_{-0.0087}^{+0.0098} \hspace{.6cm} & {\rm Planck~(TT+lowP) }
\nonumber \\
n_s =0.966 \pm 0.006\hspace{.6cm} &  {\mathrm d}n_s/{\mathrm d}\ln k =  -0.0206_{-0.0056}^{+0.0054} \hspace{.6cm}& {\rm Planck~(TT+lowP) + Ly}\alpha 
\nonumber \\
n_s = 0.966\pm0.006 \hspace{.6cm} &  {\mathrm d}n_s/{\mathrm d}\ln k =  -0.0117_{-0.0086}^{+0.0085} \hspace{.6cm} & {\rm Planck~(TT+lowP) +BKP}
\nonumber \\
n_s = 0.967\pm0.005 \hspace{.6cm} &  {\mathrm d}n_s/{\mathrm d}\ln k =  -0.0200_{-0.0053}^{+0.0054} \hspace{.6cm} & {\rm Planck~(TT+lowP) +BKP+ Ly}\alpha 
\nonumber \\
\end{eqnarray*}

\begin{figure}[htbp]
\begin{center}
\epsfig{figure= 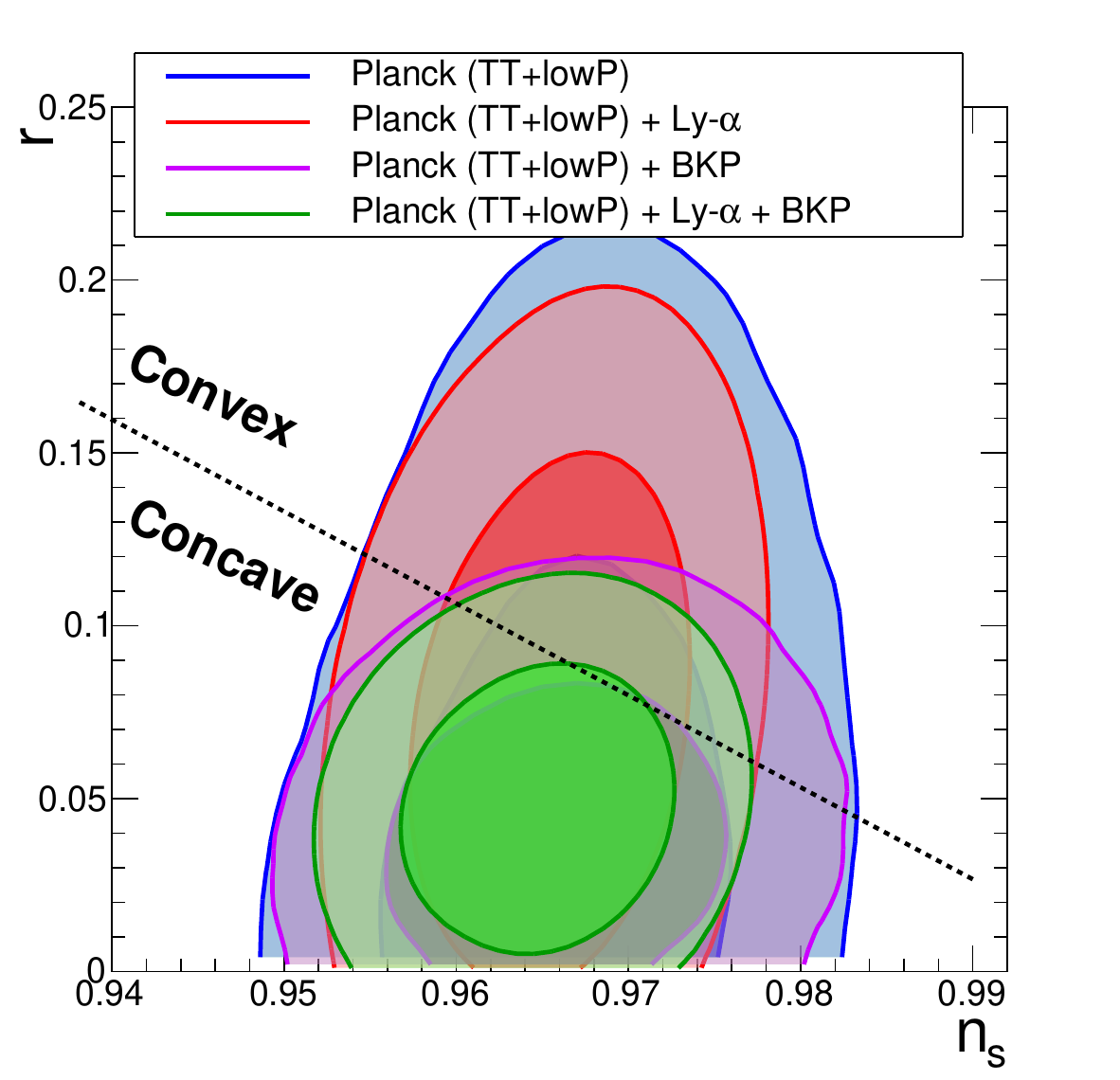,width = 5.0cm}
\epsfig{figure= 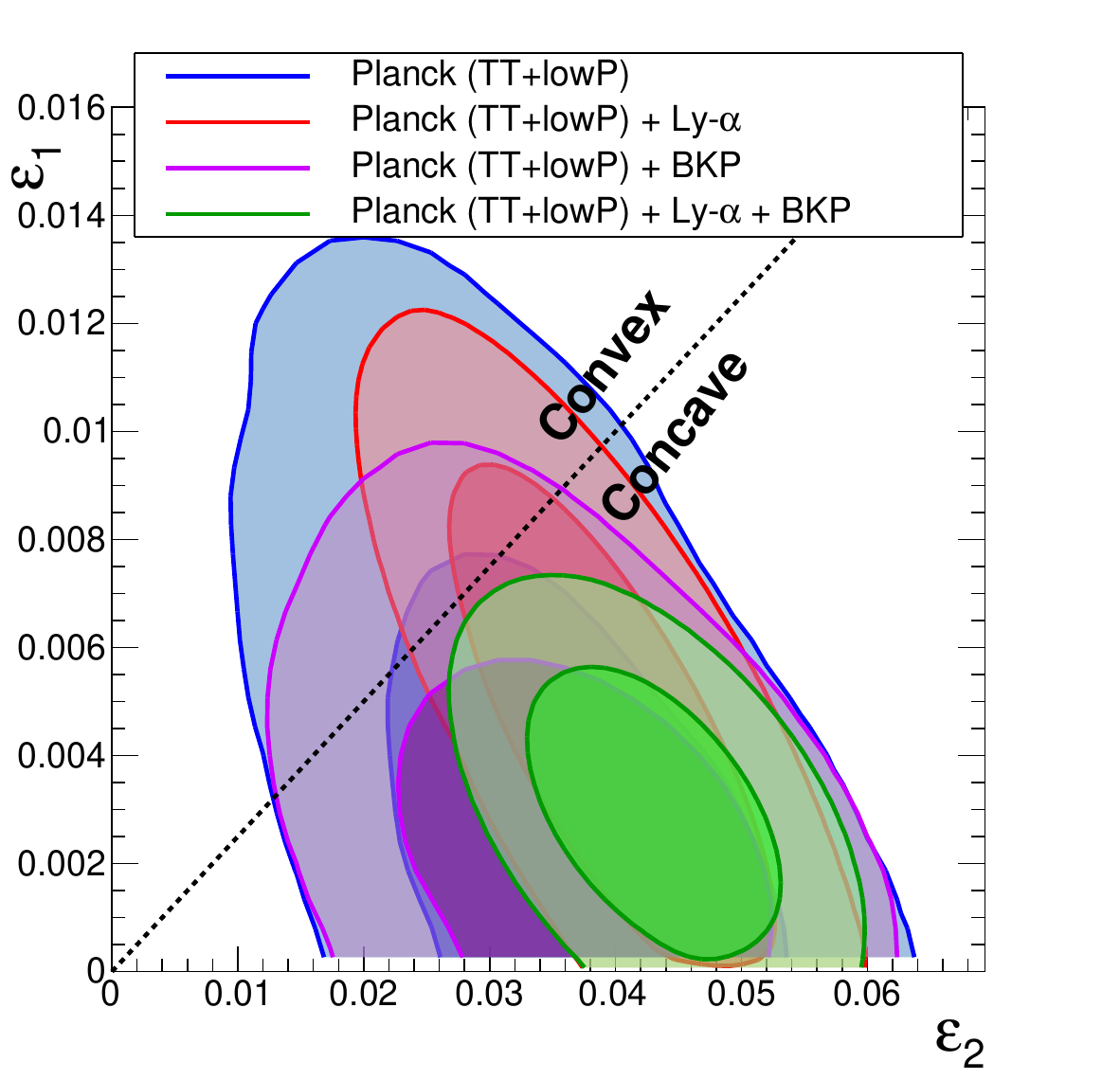,width = 5.0cm}
\epsfig{figure= 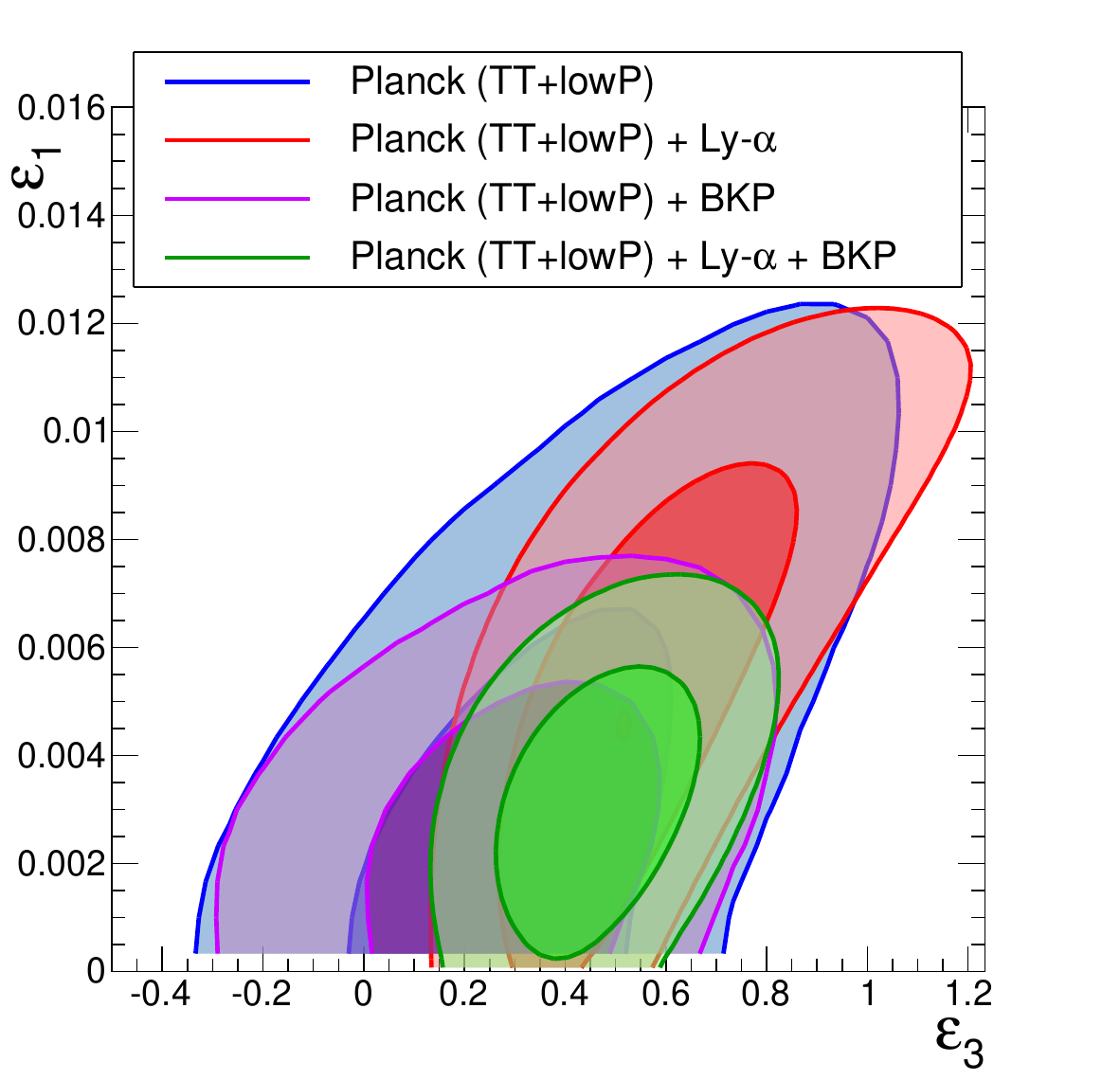,width = 5.0cm}
\caption{\it Left plot: constraints on the two parameters describing primordial fluctuations ($n_s$,$r$).  Middle plot: constraints on the two slow-roll parameters ($\varepsilon_2$,$\varepsilon_1$).  Right plot: constraints on the two slow-roll parameters ($\varepsilon_3$,$\varepsilon_1$).    The 2D 68\% and 95\% confidence contours are obtained for four combinations: Planck 2015 data  (TT+lowP),   Planck 2015 data  (TT+lowP) and the BOSS Ly$\alpha$ data ,  Planck 2015 data  (TT+lowP) and the BICEP2/Keck Array data (BKP) and finally all the data. }
\label{fig:nsrPlanck}
\end{center}
\end{figure}

The CMB and  Ly$\alpha$ measurements provide a powerful probe of cosmic inflation through the two parameters $n_s$ and ${\mathrm d}n_s/{\mathrm d}\ln k$ as  explained before. In addition,  CMB polarization makes it possible to constrain the tensor-to-scalar ratio, $r$, which is directly related to the inflation field.  Even if Ly$\alpha$ data alone cannot measure $r$,  it can be used in combination with CMB to improve the uncertainty on $r$ over what CMB alone can do, thanks to 
the correlations of $r$ with the other cosmological parameters and the tightened constraints Ly$\alpha$ provides on the latter. The improvement appears  on  the left plot of figure~\ref{fig:nsrPlanck} (blue and red curves).

The uncertainty on  $r$ can also be reduced by  direct measurement of the large-scale B-modes in CMB polarization. We therefore include the BICEP2/Keck-Array-Planck (BKP) joint analysis~\cite{BKP2015}. 
The gain provided by the B-modes for CMB data alone is visible on the left plot of figure~\ref{fig:nsrPlanck} by comparing the blue and the purple curves.  As shown in the right plot of Fig.~\ref{fig:running}, however, the BKP data has no impact on the measurement of ${\mathrm d}n_s/{\mathrm d}\ln k$. The addition of Ly$\alpha$ yields further improvement, illustrated as the green curve in both plots. We obtain an upper limit  $r<0.098$ at 95\% CL,  letting  the running  ${\mathrm d}n_s/{\mathrm d}\ln k$ free.  

The parameters of the scalar and tensor power spectra may be estimated in the framework of  slow-roll inflation (see \cite{planck2013Infla,planck2015Infla,Finelli2010}) from the value of the Hubble parameter and the hierarchy of its time derivatives. In the context of the Hubble flow-functions (HFF), these parameters are defined as: $\varepsilon_1 = - \dot{H}/H^2$, and $\varepsilon_{i+1} = - \dot{\varepsilon_i}/(H\varepsilon_i)$ with $i\geq1$.
The  scalar spectral index $n_s$,  the  tensor spectral index $n_t$ and the  running of the scalar spectral index  ${\mathrm d}n_s/{\mathrm d}\ln k$ can be related to the three slow-roll parameters $\varepsilon_1$, $\varepsilon_2$ and $\varepsilon_3$ by the following equations:
\begin{eqnarray}
n_s - 1 &  = & -2\varepsilon_1 - \varepsilon_2 - 2\varepsilon_1^2 - (2C+3)\varepsilon_1 \varepsilon_2 - C\varepsilon_2 \varepsilon_3,\\
n_t & = &  -2\varepsilon_1  - 2\varepsilon_1^2 - 2(C+1)\varepsilon_1 \varepsilon_2,\\
{\mathrm d}n_s/{\mathrm d}\ln k & = &  - 2\varepsilon_1 \varepsilon_2 - \varepsilon_2 \varepsilon_3, 
\label{eq:slow-roll-predictions}
\end{eqnarray}
where $C\simeq -0.7296$. The tensor-to-scalar power ratio $r$ can be derived from the tensor index $n_t$ through the consistency relation $n_t  =  -r(2-r/8-n_s)/8$, obtained  at  second order when inflation is driven by a single slow-rolling scalar field. 

 The middle plot of figure~\ref{fig:nsrPlanck} shows the 2D contours for two slow-roll parameters $\varepsilon_1$ and $\varepsilon_2$,  letting free the running index $\varepsilon_3$. This approach is  equivalent, in frequentist interpretation,  to marginalizing over $\varepsilon_3$ in Bayesian analysis. The constraints on these two slow-roll parameters are given below  for four different data sets:
\begin{eqnarray*}
\varepsilon_1<0.0126\,(95\%{\rm CL}) \hspace{1cm} &  \varepsilon_2 =  0.0343 ^{+0.0104}_{-0.0104} \hspace{1cm} & {\rm Planck~(TT+lowP) }\\
\varepsilon_1 <0.0108 \,(95\%{\rm CL}) \hspace{1cm} &  \varepsilon_2 =  0.0393 ^{+0.0089}_{-0.0083} \hspace{1cm} & {\rm Planck~(TT+lowP) + Ly}\alpha \\
\varepsilon_1 <0.0067 \,(95\%{\rm CL}) \hspace{1cm} &  \varepsilon_2 =  0.0365 ^{+0.0095}_{-0.0093} \hspace{1cm} & {\rm Planck~(TT+lowP) + BKP}\\
\varepsilon_1 <0.0063\,(95\%{\rm CL}) \hspace{1cm} &  \varepsilon_2 =  0.0431 ^{+0.0065}_{-0.0064} \hspace{1cm} & {\rm Planck~(TT+lowP) + BKP + Ly}\alpha \\
\end{eqnarray*}

The third parameter in the HFF formalism,  $\varepsilon_3$, is constrained by the measurement of ${\mathrm d}n_s/{\mathrm d}\ln k$.  As shown on the right plot of figure~\ref{fig:nsrPlanck},  Ly$\alpha$  data improves the determination of $\varepsilon_3$. Its positive best-fit value directly reflects the non-zero detection of running.  
 We should stress that a large running leads to values of $\varepsilon_3$ much larger than those of $\varepsilon_1$, $\varepsilon_2$, suggesting that the HFF expansion might not be convergent, and that the use of slow-roll expressions truncated at order three like in Eq.~(\ref{eq:slow-roll-predictions}) is not necessarily appropriate in this context.
In a future study,  we will develop an analysis similar to  the one done in~\cite{Lesgourgues2007,Lesgourgues2008,planck2015Infla}  to  directly measure the  shape of the inflation potential that provides a good fit to the data, without requiring any kind of slow-roll expansion. 

\subsection{Discussion on $\sum m_\nu$ and ${\mathrm d}n_s/{\mathrm d}\ln k$}
\label{sec:discussion}
The joint Ly$\alpha$ + CMB analysis presented in this paper allows us to set stringent constraints on cosmology, providing significant improvements upon CMB  alone  from \citet{Planck2015} on two main fronts: the constraint on $\sum m_\nu$ is much tighter, and a running of $n_s$ is measured at more than $3~\sigma$. We here discuss these two results in terms of their robustness and correlations.

The constraint on $\sum m_\nu$, on the one hand, comes from the measurement of $\sigma_8$ in Ly$\alpha$ data, and from the correlation between $\sigma_8$ and $\sum m_\nu$ provided by CMB. The value of $\sigma_8$ is derived from the normalization of the 1D flux power spectrum, and thus from its measurement at zeroth order. 

The value of ${\mathrm d}n_s/{\mathrm d}\ln k$, on the other hand, is derived from the different values of $n_s$ determined by CMB and Ly$\alpha$ (e.g. from the measurement of the 1D flux power spectrum at first order to access the slope information that determines $n_s$), and from variations of $n_s$ within the scales probed by either  probe (e.g. from the measurement of the 1D flux power spectrum at second order). It is therefore more sensitive to systematic effects  in the measurement of the flux power spectrum. 
For instance, the slope of the 1D flux power spectrum is sensitive to  the modeling of  instrumental effects (such as spectrograph resolution) as well as of physical contributions affecting the intergalactic medium (such as SN or AGN feedback, UV fluctuations, etc.).  These effects have been included in this work through nuisance parameters that are fitted along with the relevant cosmological parameters. This is nevertheless a delicate task, not free of any possible yet-unaccounted-for additional systematic which could affect the determination of $n_s$. 

We can estimate 
 the value of  ${\mathrm d}n_s/{\mathrm d}\ln k$  from the difference of scale factors at the CMB and Ly$\alpha$ pivot scales, $k_{\rm CMB} = 0.05~{\rm Mpc}^{-1}$ and $k_{Ly\alpha} \sim  0.7~{\rm Mpc}^{-1}$, respectively. Given the definition of running of Eq. ~\ref{eq:running}, and the values of $n_s$ determined separately from CMB or Ly$\alpha$ data, we estimate ${\mathrm d}n_s/{\mathrm d}\ln k$ to be approximately -0.02. This is in  agreement with the best-fit value $ -0.0178_{-0.0048}^{+0.0054} $ (cf. Sec.~\ref{sec:primordial}), thus confirming that running is indeed detected in this work mostly from the different levels of $n_s$ in CMB and Ly$\alpha$ data, and thus mostly from a first order measurement of the Ly$\alpha$ power spectrum.  Any unidentified systematic uncertainty that would resolve the tension on $n_s$ would thus simultaneously annihilate our detection of ${\mathrm d}n_s/{\mathrm d}\ln k$. 
 
Finally, let us study the  impact on $\sum m_\nu$ of the $2.3~\sigma$ tension on $n_s$. In Paper~I (Sec. 4.2.2), we showed that it did not affect the result since the tension on $n_s$ is roughly the same for small (0.1 eV) or large (0.3 eV) neutrino masses. In the present work, we infer a similar conclusion from the fact that $n_s$ and $\sum m_\nu$ have little correlation, both in Ly$\alpha$ and in CMB data (cf. Sec. \ref{sec:LyaAlone} and \ref{sec:LyaCMB}). As a final test, we implemented a dedicated MCMC for CMB data, allowing both $\sum m_\nu$ and ${\mathrm d}n_s/{\mathrm d}\ln k$ to vary. The result, given in Sec.~\ref{sec:primordial}, indicates a null correlation between these two parameters. Errors or limits on both parameters are also in excellent agreement with the values obtained when only one of them at a time is included in the fit, confirming the absence of degeneracy between $\sum m_\nu$ and ${\mathrm d}n_s/{\mathrm d}\ln k$. 
In conclusion, if in the future a systematic effect is found that affects the determination of $n_s$, it would directly alter the determination of ${\mathrm d}n_s/{\mathrm d}\ln k$ but would not modify significantly  the limit on $\sum m_\nu$.


\section{Conclusions}
\label{sec:conclusion}
In  this paper, we present  an update on the constraints we derive on several cosmological parameters using Ly$\alpha$ data, either taken alone or in combination with  CMB and BAO data. We  improve upon our previous study of Paper~I \cite{Palanque2015a} on several fronts that we summarize below.

 We  improved the likelihood describing the Ly$\alpha$ data in several ways. We  relaxed our model of the IGM, now modeling the redshift-dependence of the IGM temperature $T_0$ and its $\delta$ dependence (i.e., the logarithmic slope $\gamma$) as, respectively, a broken and a single power law.  The likelihood now includes additional freedom to account for the systematic uncertainties that we  identified. We ran new hydrodynamical simulations that allow us to improve our model  of the impact of the splicing technique on the 1D power spectrum. This  splicing technique is  used in the simulation grid to mimic large high-resolution simulations equivalent to $3092^3$ particles per species in a $(100~h^{-1}~{\rm Mpc})^3$ box.  We also better estimate the contribution of sample variance to the simulation uncertainties. Using this updated likelihood, we set a robust upper bound on the sum of the neutrino masses $\sum m_\nu < 1.1$~eV (95\% C.L.) from Ly$\alpha$ data alone, prioritizing over all known systematics.  
Compared to the  Planck 2015 measurement~\cite{Planck2015}, the constraints we derive on cosmological parameters $\Omega_m$ and $\sigma_8$ are in excellent agreement, but we note    a 2.3~$\sigma$ discrepancy on $n_s$. 
 
 In a second step, we combine our new Ly$\alpha$ likelihood with the MCMC chains from Planck 2015. In the context of a flat $\Lambda$CDM$\nu$ cosmology, this combined study leads to the tightest limit published   on $\sum m_\nu$, with an upper bound of 0.12~eV (95\% C.L.) using Ly$\alpha$ and Planck TT+lowP data. The addition of BAO does not further improve this limit. 
 
 In a third step, we investigate the improvement, over CMB data alone,  that Ly$\alpha$ data can provide  on other parameters. In particular, we show that  through the correlation between the reionization optical depth $\tau$ and $\sigma_8$, Ly$\alpha$ data can reduce the uncertainties on the measurement of $\tau$. The central value remains fully compatible with the one derived from CMB.  
 
 Finally, we consider extensions to the $\Lambda$CDM$\nu$ cosmology, in particular focusing on a possible running of the scalar spectral index. We note a clear improvement of the total $\chi^2$  when allowing for running, in agreement with the observed tension on $n_s$ between Planck and Ly$\alpha$ data sets. However, this improvement, and the subsequent $3~\sigma$ detection of running, could result from a coincidence between a
 $\sim 1~\sigma$ effect in Planck (due to the mismatch between the high and low multipoles in the temperature power spectrum) on the one hand, and a $\sim 2.3~\sigma$ effect in Ly$\alpha$ data (possibly coming from an unidentified systematic effect) on the other hand. Allowing for ${\mathrm d}n_s/{\mathrm d}\ln k$ to vary in the fit does not change our best limit on $\sum m_\nu$, but in that case BAO data are needed to reach that best limit. The combination of Ly$\alpha$ with Planck (TT, TE, EE + lowP) and BAO still leads to $\sum m_\nu < 0.12$~eV at 95\% C.L.,  even when allowing for running. 

As an interesting consequence of our precise measurement of ${\mathrm d}n_s/{\mathrm d}\ln k$, we can derive constraints on the inflation potential. In the context of slow-roll inflation, we use the second-order expansion in Hubble flow-functions  to relate the  
slow-roll parameters $\epsilon_1$, $\epsilon_2$ and $\epsilon_3$ to the power spectrum parameters $n_s$, $n_t$, and ${\mathrm d}n_s/{\mathrm d}\ln k$. The uncertainties on the inflation parameters are reduced significantly compared to their measurement with CMB data.

\acknowledgments

We acknowledge PRACE (Partnership for Advanced Computing in Europe) for awarding us access to resources Curie thin nodes and Curie fat nodes, based in France at TGCC.\\
This work was also granted access to the resources of CCRT under the allocation 2013-t2013047004 made by
GENCI (Grand Equipement National de Calcul Intensif).\\
N.P.-D., Ch.Y. and G.R.  acknowledge  support from grant ANR-11-JS04-011-01 of Agence Nationale de la Recherche.\\
M.V. is supported by ERC-StG "CosmoIGM".\\
The work of G.R. is also supported by the National Research Foundation of Korea (NRF) through NRF-SGER 2014055950 funded by the Korea government (MOE), and by the faculty research fund of Sejong University in 2014.\\
We thank Volker Springel for making \texttt{GADGET-3} available to our team.


\bibliographystyle{unsrtnat_arxiv}
\bibliography{biblio}

\end{document}